\documentclass[a4paper,11pt]{article}
\pdfoutput=1 

\usepackage{jheppub} 
\usepackage{amsmath,amssymb,amsthm,amscd}
\input epsf.sty
\usepackage{subfig}
\usepackage{graphicx}
\usepackage[export]{adjustbox}
\usepackage[table]{xcolor}                  
\usepackage{colortbl}
\usepackage{booktabs}                       
\usepackage{multirow,bigdelim}              
\usepackage{longtable}                      
\usepackage{units}                          
\usepackage{slashed}                        
\usepackage{tikz}                           
\usepackage{float}
\usepackage[export]{adjustbox}              
\usepackage{url}                            

\theoremstyle{definition}

\newcommand{\CC}{{\cal C}}

\newcommand{\CI}{{\cal I}}

\newcommand{\CN}{{\cal N}}
\newcommand{\CO}{{\cal O}}

\newcommand{\CW}{{\cal W}}

\newcommand{\CY}{{\cal Y}}
\newcommand{\CZ}{{\cal Z}}

\def\IZ{{\mathbb Z}}
\def\IR{{\mathbb R}}
\def\IC{{\mathbb C}}

\def\IS{{\mathbb S}}
\def\IF{{\mathbb F}}

\newcommand{\re}{{\rm e}}
\newcommand{\ri}{{\rm i}}
\newcommand{\rd}{{\rm d}}

\newcommand{\Tr}{{\rm Tr}}

\newcommand{\mO}{\mathsf{O}}

\newcommand{\mx}{\mathsf{x}}

\newcommand{\mm}{\mathsf{p}}

\newcommand{\mH}{\mathsf{H}}
\newcommand{\mJ}{\mathsf{J}}

\newcommand{\bs}{\boldsymbol}
\newcommand{\balpha}{\boldsymbol{\alpha}}
\newcommand{\blam}{\boldsymbol{\lambda}}

\usepackage{color}

\newcommand{\be}{\begin{equation}}
\newcommand{\ee}{\end{equation}}
\newcommand{\ba}{\begin{aligned}}
\newcommand{\ea}{\end{aligned}}
\newcommand{\ben}{\begin{eqnarray}\displaystyle}
\newcommand{\een}{\end{eqnarray}}

\def\mc{\mathcal}
\def\md{\mathbf}

\def\IC{\mathbb{C}}

\def\IF{\mathbb{F}}

\def\IR{\mathbb{R}}
\def\IZ{\mathbb{Z}}

\def\({\left(}
\def\){\right)}
\def\[{\left[}
\def\]{\right]}

\newcommand{\pd}{\partial}

\newcommand{\nn}{\nonumber \\}

\newdimen\tableauside\tableauside=1.0ex
\newdimen\tableaurule\tableaurule=0.4pt
\newdimen\tableaustep
\def\phantomhrule#1{\hbox{\vbox to0pt{\hrule height\tableaurule width#1\vss}}}
\def\phantomvrule#1{\vbox{\hbox to0pt{\vrule width\tableaurule height#1\hss}}}
\def\sqr{\vbox{%
  \phantomhrule\tableaustep
  \hbox{\phantomvrule\tableaustep\kern\tableaustep\phantomvrule\tableaustep}%
  \hbox{\vbox{\phantomhrule\tableauside}\kern-\tableaurule}}}
\def\squares#1{\hbox{\count0=#1\noindent\loop\sqr
  \advance\count0 by-1 \ifnum\count0>0\repeat}}
\def\tableau#1{\vcenter{\offinterlineskip
  \tableaustep=\tableauside\advance\tableaustep by-\tableaurule
  \kern\normallineskip\hbox
    {\kern\normallineskip\vbox
      {\gettableau#1 0 }%
     \kern\normallineskip\kern\tableaurule}%
  \kern\normallineskip\kern\tableaurule}}
\def\gettableau#1{\ifnum#1=0\let\next=\null\else
\squares{#1}\let\next=\gettableau\fi\next}

\title{\huge{Non-perturbative approaches to the quantum Seiberg--Witten curve}}

\author{Alba Grassi$^a$, Jie Gu$^b$ and Marcos Mari\~no$^b$}

\affiliation{
$^a$Simons Center for Geometry and Physics,\\
 SUNY, Stony Brook, NY, 1194-3636, USA \\
 \\
 $^b$ D\'epartement de Physique Th\'eorique et Section de Math\'ematiques,\\
Universit\'e de Gen\`eve, Gen\`eve, CH-1211 Switzerland}

\emailAdd{agrassi@scgp.stonybrook.edu, jie.gu@unige.ch, marcos.marino@unige.ch}

\abstract{We study various non-perturbative approaches to the
  quantization of the Seiberg--Witten curve of ${\cal N}=2$, $SU(2)$
  super Yang--Mills theory, which is closely related to the modified
  Mathieu operator. The first approach is based on the quantum WKB
  periods and their resurgent properties. We show that these
  properties are encoded in the TBA equations of Gaiotto--Moore--Neitzke determined by the
  BPS spectrum of the theory, and we relate the
  Borel-resummed quantum periods to instanton calculus. In addition, we
  use the TS/ST correspondence to obtain a closed formula for the
  Fredholm determinant of the modified Mathieu operator.  
 Finally, by using blowup equations, we explain the connection between this 
 operator and the $\tau$ function of Painlev\'e $\rm III$.}

\begin{document}
\maketitle

\flushbottom

\section{Introduction}

In recent years, many interesting and surprising relations have been
obtained between quantum mechanical systems, on one hand, and
supersymmetric gauge theories and topological strings, on the other
hand. One example of such a relation is the gauge/Bethe correspondence
of \cite{ns}, which connects quantum integrable systems to instanton
calculus in gauge theory. A second example is the 
topological string/spectral theory (TS/ST) correspondence, which
provides explicit predictions for the spectral determinants of quantum
mirror curves \cite{ghm,cgm,mmrev}. Finally, the study of BPS states
in supersymmetric gauge theories turns out to be closely related to
the WKB method as applied to Seiberg--Witten (SW) curves
\cite{gmn,gmn2,oper}. This relation can be upgraded to include
resurgent properties of the quantum periods \cite{ims, mm-s2019}.  All
these connections can be used to obtain new results in quantum theory
from gauge/string theory. For example, the results of \cite{ns,ghm}
lead to new exact quantization conditions for the spectrum of the
relevant operators. Conversely, one can use quantum mechanical results
to derive new results of string/gauge theories, like for example
non-perturbative definitions of topological string partition functions
on local Calabi--Yau (CY) manifolds \cite{ghm,cgm,mz}.

Perhaps the simplest quantum-mechanical model where all these methods
can be applied is the quantum version of the SW curve for
${\cal N}=2$, $SU(2)$ super Yang--Mills (SYM) theory.  The
corresponding operator is the (modified) Mathieu operator, which is a
traditional chapter in the theory of Schr\"odinger operators. This
operator has been also revisited in the context of supersymmetric
gauge theory and topological string theory in various works (see
e.g. \cite{mirmor,he-miao,huangNS,basar-dunne,kpt,ashok,coms}), but
many important aspects have not been discussed yet.  In this paper we
use methods from supersymmetric gauge theory and topological string
theory to obtain quantum-mechanical properties of the modified Mathieu
operator at the non-perturbative level, and we test these properties
against first-principles computations. We also discuss the
relationships between these different approaches.

The first aspect that we explore is the resurgent structure of the
quantum periods, which we review in section \ref{wkb}. Building on
\cite{gmn2}, Gaiotto considered in \cite{oper} the conformal limit of
the TBA equations of \cite{gmn} for an $\CN=2$ supersymmetric gauge
theory, and he conjectured that the resulting integral equations
describe the quantum periods for the corresponding quantum SW
curve. In the case of Argyres--Douglas theories, this problem was
studied in detail in \cite{ims}, which pointed out precise connections
to the resurgent properties of these periods, and used these
properties to derive the conjecture of \cite{oper} in the case of
general polynomial potentials

In section \ref{tbagmn} of this paper we use the conformal limit of
the TBA equations to obtain a prediction for these
resurgent properties in the case of the modified Mathieu operator. In
particular, we obtain the precise structure of the Stokes
discontinuities of the quantum periods.  We then test these predictions against
first-principles calculations in the all-orders WKB method, in
particular against high order results for the expansion of the quantum
periods. We also comment on how to use these TBA equations to compute
Borel resummations of the quantum periods.

As pointed out in \cite{mirmor} and explored in many subsequent
papers, the NS limit of instanton calculus \cite{ns} provides a
different resummation of the WKB expansion, in terms of a convergent
expansion in the instanton counting parameter. However, this
resummation has a very different flavor from the Borel resummation
appearing in the theory of resurgence, and it is important to have a
precise dictionary between the two types of resummation.  We address
this issue in section \ref{res}.

As we mentioned above, the TS/ST correspondence gives explicit
expressions for spectral determinants of operators obtained in the
quantization of mirror curves.  As pointed out in \cite{gm3}, there is
a four-dimensional limit of the correspondence in which the relevant
operator is the quantization of the SW curve for pure ${\cal N}=2$,
$SU(N)$ Yang--Mills theory. This leads to a spectral problem which is
different from the one considered in \cite{ns} for $N>2$. In the case
of the $SU(2)$ theory considered in this paper, the spectral problems
coincide, but the TS/ST correspondence gives, in addition to the
quantization condition of \cite{ns}, an explicit expression for the
spectral determinant, which we derive in detail in section \ref{tssst}
of this paper.  The resummed quantum periods defined by instanton
calculus are key ingredients in this expression. We test the resulting
formula and in particular we compare our result to the TBA equation
describing this spectral determinant which was conjectured by
Al. B. Zamolodchikov in \cite{post-zamo}.

 In section \ref{mathieu-painleve}, based on previous
works, we use the vanishing Nakajima-Yoshioka blowup equations to
prove that the exact spectrum of the modified Mathieu operator is computed by the
zeros of the $\tau$ function of Painlev\' e $\rm III_3$.  Finally, in
section \ref{concl} we conclude and discuss some open problems.

We have also included two Appendices: in the first one we extend the
derivation of section \ref{tssst} to $SU(N)$ quantum SW curves, while
in the second one we review some of the results of Zamolodchikov's
paper \cite{post-zamo}.

\section{The all-orders WKB method}\label{wkb}

Our first approach to the quantum SW curve will be based on the
so-called exact WKB method, see for example
\cite{voros,voros-quartic,ddpham,in-exactwkb}.  We will now summarize
the basic ingredients of the theory.

The Schr\"odinger equation for a non-relativistic particle in a
potential $V(x)$ and with energy $E$ reads as follows:
\be
\label{schrodinger}
-\hbar^2 \psi''(x) + (V(x)-E) \psi(x)=0. 
\ee
The standard WKB method produces asymptotic expansions in $\hbar$ for
the solutions to this equation.  Let us consider the following ansatz
for the wavefunction,
\begin{equation}\label{eq:psi-WKB}
  \psi(x) = \exp\(\frac{\ri}{\hbar}\int^x Y(x',E;\hbar) \rd x'\) \ .
\end{equation}
The function $Y(x,E;\hbar)$ satisfies the Riccati equation
\begin{equation}\label{riccati}
  Y^2 - \ri \hbar \frac{\rd Y}{\rd x} = E - V(x)
  \ .
\end{equation}
It has the formal power series expansion in powers of $\hbar$
%
\begin{equation}\label{ricwkb}
  Y(x,E;\hbar) = \sum_{n=0}^\infty p_n(x,E) \hbar^n \ ,
\end{equation}
where in particular $p_0(x, E)$ is the classical momentum as a
function of $x$ and the conserved energy. If one splits $Y$ into the even component and the odd component,
\be
Y = p_{\text{even}}+p_{\text{odd}}, 
\ee
 with
\begin{equation}
\label{p-series}
  p_{\text{even}}(x,E;\hbar) = \sum_{n=0}^\infty p_{2n}(x,E)\hbar^{2n}
  \ ,\quad
  p_{\text{odd}}(x,E;\hbar) = \sum_{n=0}^\infty
  p_{2n+1}(x,E)\hbar^{2n+1} \ ,
\end{equation}
one finds that the odd component is in fact a total derivative
\begin{equation}
  p_{\text{odd}}(x,E;\hbar) = \frac{\ri\hbar}{2}\frac{\rd}{\rd x} \log
  p_{\text{even}}(x,E;\hbar).
\end{equation}
By substituting \eqref{eq:psi-WKB} into the Schr\"{o}dinger equation,
one finds (see for instance \cite{bpv})
\begin{gather}
  p_{2n} = (-1)^n v_{2n} \ , \quad n \geq 0 \\
  v_{n} = \frac{1}{2p_0}\(\pd_xv_{n-1} - \sum_{k=1}^{n-1}v_k v_{n-k}\) \ ,
\end{gather}
from which the components $p_{2n}(x,E)$ can be solved recursively,
starting from the known expression of $p_0$.

Geometrically, we can regard $p_{\rm even}(x, E;\hbar) \rd x$ as a
meromorphic differential on the curve defined by
\be
\label{wkb-curve} 
y^2 = 2(E- V(x)). 
\ee
We will call it the {\it WKB curve}, and we will denote it as
$\Sigma_{\rm WKB}$.  This curve depends on a set of moduli which
include the energy $E$ and the parameters of the potential $V(x)$. The
basic objects in the exact WKB method are the periods of
$p_{\rm even} (x, E;\hbar) \rd x$ along one-cycles of
$\Sigma_{\rm WKB}$, which we will call {\it WKB periods} or {\it
  quantum periods}. We will denote them as
\be
\Pi_\gamma(\hbar)= \oint_\gamma p_{\rm even} (x,E;\hbar) \rd x, \qquad \gamma \in H_1(\Sigma_{\rm WKB}), 
\ee
and they are formal power series in even powers of $\hbar$, just like
$p_{\rm even}(x)$,
\be
\label{pi-wkb-expansion}
\Pi_\gamma(\hbar)=\sum_{n \ge 0} \Pi^{(n)}_\gamma \hbar^{2n}, \qquad
\Pi^{(n)}_\gamma = \oint_\gamma p_n(x, E) \rd x.  
\ee
Note that the coefficients $ \Pi^{(n)}_\gamma$ depend on the moduli of
the WKB curve. We will call $\Pi^{(0)}_\gamma$ the {\it classical
  periods}. The calculation of these coefficients at high order can be
quite involved, even for simple quantum systems.

In this paper we are interested in the modified Mathieu Hamiltonian,
with the conventions
\begin{equation}
  H(p,x) =p^2+V(x), \qquad  V(x)= 2\Lambda^2 \cosh x.
\label{eq:su2curve}
\end{equation}
Upon quantization, we obtain the operator 
\be
\label{eq:schrodinger}
\mH= \mm^2+ 2 \Lambda^2 \cosh(\mx), \qquad [\mx, \mm]= \ri \hbar. 
\ee
We will refer to this as the modified Mathieu operator. It is
well-known that the WKB curve of the modified Mathieu Hamiltonian
happens to coincide with the SW curve of ${\cal N}=2$, $SU(2)$
Yang--Mills theory, in the conventions appropriate for the relation to
integrable systems (see e.g. \cite{lerche-rev} for a review of SW
theory and \cite{swbook} for its connection to integrable
systems). In order to do this, we identify $E$ with the Coulomb
modulus $u$ by
\begin{equation}
  E = 2u \ .
\end{equation}

Let us first consider the classical periods of the modified Mathieu
equation. Since the WKB curve is a torus, there will be two periods,
corresponding to the two cycles of the torus.  The $B$ period
corresponds to the classical volume of phase space
\begin{equation}\label{eq:PiB0}
  \Pi_B^{(0)}(E)=4\,\ri\int_0^{x_+} \rd x \, \sqrt{E-2\Lambda^2 \cosh
    x},\ee 
  where 
  \be x_+=\cosh^{-1} \frac{E}{2\Lambda^2}
\end{equation}
is the turning point. This classical period can be evaluated
explicitly as
\begin{equation}\label{PB0}
  \Pi_B^{(0)}(E)=8\,\ri\sqrt{E+2\Lambda^2} \left[ {\bf K}\left(
      \frac{E-2\Lambda^2}{E+2\Lambda^2} \right) 
    -{\bf E} \left( \frac{E-2\Lambda^2}{E+2\Lambda^2} \right) \right].
\end{equation}
(We denote the elliptic integrals with boldface letters ${\bf K}$,
${\bf E}$, and their argument is the squared modulus $m=k^2$).  There
is in addition an $A$ period which corresponds to motion along the
imaginary axis. Classically, it is given by,
\begin{equation}\label{eq:PiA0}
  \Pi_A^{(0)}(E)
  = 
  -2\,\ri
  \int_{-\pi \ri }^{\pi \ri } \rd x \,
  \left(  \sqrt{E-2\Lambda^2 \cosh x}  \right) 
  =8 {\sqrt{E+2\Lambda^2}}  {\bf E} \left( {4
      \Lambda^2 \over 2 \Lambda^2+E} \right).  
\end{equation}
In the simplest case when $E=0$ and $\Lambda=1$, we have
%
\begin{equation}
  \Pi_A^{(0)}(0) = (1+\ri) \frac{16\pi^{3/2}}{\Gamma(1/4)^2},\quad
  \Pi_B^{(0)}(0) = -\ri \frac{16\pi^{3/2}}{\Gamma(1/4)^2} .
\end{equation}
We note that these classical periods are, up to normalization, the
famous $a$ and $a_D=\partial_a F$ periods of SW theory \cite{sw},
namely
\be
\label{abcpe}  
\Pi^{(0)}_A(E)=2 \pi a(u), \qquad \Pi_B^{(0)}(E) ={ 2 \ri }a_D(u). 
\ee
We will denote the all-orders WKB quantum periods as
\begin{equation}
  \label{QP}  \Pi_{A, B} (E,\hbar) =\sum_{n=0}^\infty \hbar^{2n} \Pi_{A, B}^{(n)} (E). 
\end{equation}

In the case of the modified Mathieu equation, the most efficient way
to calculate the quantum corrections is the so-called quantum operator
approach (see e.g. \cite{huangNS}). It turns out that, for each
function $p_{2n}(x, E)$ appearing in (\ref{p-series}), one can find a
first order differential operator $\mc O_n(E)$ such that
\begin{equation}
  \mc O_n(E)\circ p_0(x,E) = p_{2n}(x,E)
\end{equation}
up to a total derivative.
Since $\mc O_n(E)$ commutes with integration, one immediately has
\begin{equation}
  \Pi_{A,B}^{(n)}(E) = \mc O_n(E)\circ \Pi_{A,B}^{(0)}(E) \ .
\end{equation}
In this way, we have computed quantum corrections up to order 193.  As
a simple example, with $\Lambda=1$ we have \cite{huangNS}
\begin{equation}
  \mc O_1 (E) = \frac{E}{48(4-E^2)} + \frac{\pd}{24\pd E} \ .
\end{equation}
Therefore, 
\begin{equation}
  \Pi_A^{(1)}(E=0) = -\frac{1-\ri}{6\sqrt{2}}\md K(-1)\ ,\quad
  \Pi_B^{(1)}(E=0) = -\frac{\ri}{6\sqrt{2}}\md K(-1) \ .
\end{equation}
We recall that the quantum periods satisfy the so-called quantum
Matone relation
\cite{matone,francisco,basar-dunne,bdu-quantum,gorsky,coms}. One of
the consequences of this relation is that
\begin{equation}\label{eq:qMatone}
  \Pi^{(0)}_A(E) \Pi^{(1)}_B(E) - \Pi^{(0)}_B(E) \Pi^{(1)}_A(E) =
  \text{const.} \ , 
\end{equation}
which we can then evaluate at $E=0$ to be
$-2 \pi \ri /3$. 



It is well-known that the formal power series appearing in the quantum
periods diverge generically as \cite{bpv,cm-ha}
\begin{equation}
  \Pi_\gamma^{(n)}\approx (2n)!, \quad n \gg 1.
\end{equation}
Therefore the expressions \eqref{pi-wkb-expansion} are just formal
power series and need to be properly resummed. A natural way of doing
so is to perform the Borel resummation. In general, given an
asymptotic series of the form
\begin{equation}
  F = \sum_{n=0}^\infty f_n \hbar^{2n} \ ,\quad\hbar\in\IC \ ,
\end{equation}
with
\begin{equation}
  f_n \sim (2n)!\ , \quad n\gg 1
\end{equation}
we split $\hbar = \re^{\ri\phi}|\hbar|$, and define the Borel
resummation to be
\begin{equation}\label{eq:sF}
  s(F)(\hbar) =
  \frac{1}{|\hbar|} \int_0^\infty \widehat F(\re^{\ri\phi}\zeta)
  \re^{-\zeta/|\hbar|} \rd \zeta\ ,
\end{equation}
where $\widehat  F (\zeta)$ is the Borel transform
\begin{equation}
  \widehat F (\zeta) = \sum_{n=0}^\infty \frac{f_n}{(2n)!} \zeta^{2n} \ .
\end{equation}
The analytic properties of $\widehat F(\zeta)$ in the $\zeta$-plane,
also called the Borel plane, are crucial. If the Borel transform has
singularities along the ray $\arg(\zeta) = \phi$, the series
$F(\hbar)$ is not Borel summable, as the integral in the Laplace
transformation \eqref{eq:sF} is obstructed. We can however deform
slightly the integration contour below or above the positive real
axis, obtaining in this way the so-called \emph{lateral Borel
  resummations} of the formal power series $F(\hbar)$:
\begin{equation}
  s_{\pm}(F)(\hbar) =
  \frac{1}{|\hbar|}
  \int_{0}^{\re^{ \ri 0^{\pm}} \infty } \widehat F (\re^{\ri\phi}\zeta)
  \re^{-\zeta/|\hbar|} \rd \zeta .
\end{equation}
These lateral resummations are in general
different, and their difference is defined as the {\it Stokes
  discontinuity} of $F$:
\be
\label{stokes-disc}
{\rm disc}(F)(\hbar) = s_+(F)(\hbar)- s_-(F)(\hbar). 
\ee
Stokes discontinuities play a crucial r\^ole in the theory of
resurgence, see e.g. \cite{abs}.

\begin{figure}
  \centering
  \subfloat[$\Pi_A$]{\includegraphics[width=0.4\linewidth,valign=c]
    {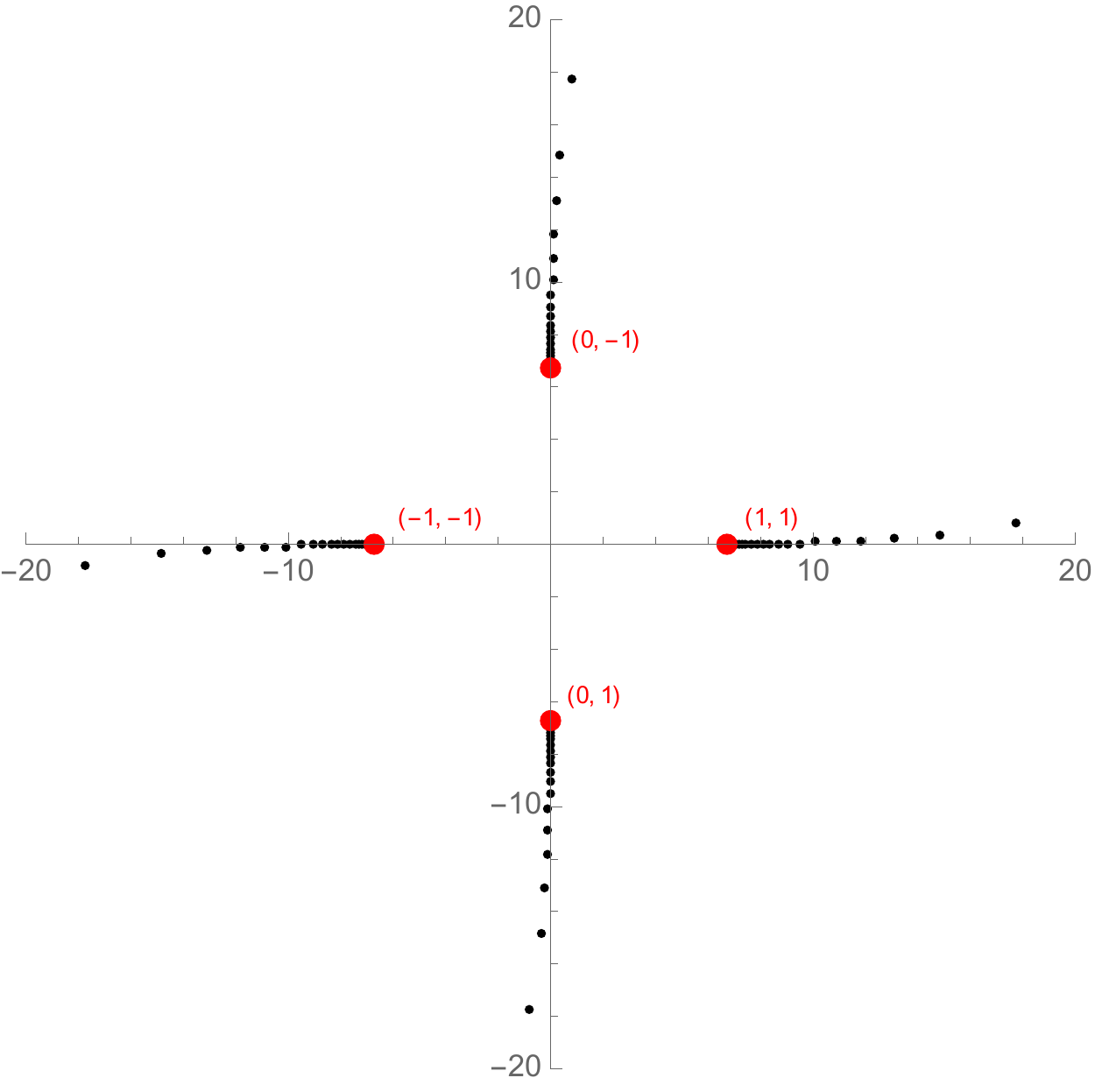}} 
  \hspace{4ex}
  \subfloat[$\Pi_B$]{\includegraphics[width=0.4\linewidth,valign=c]
    {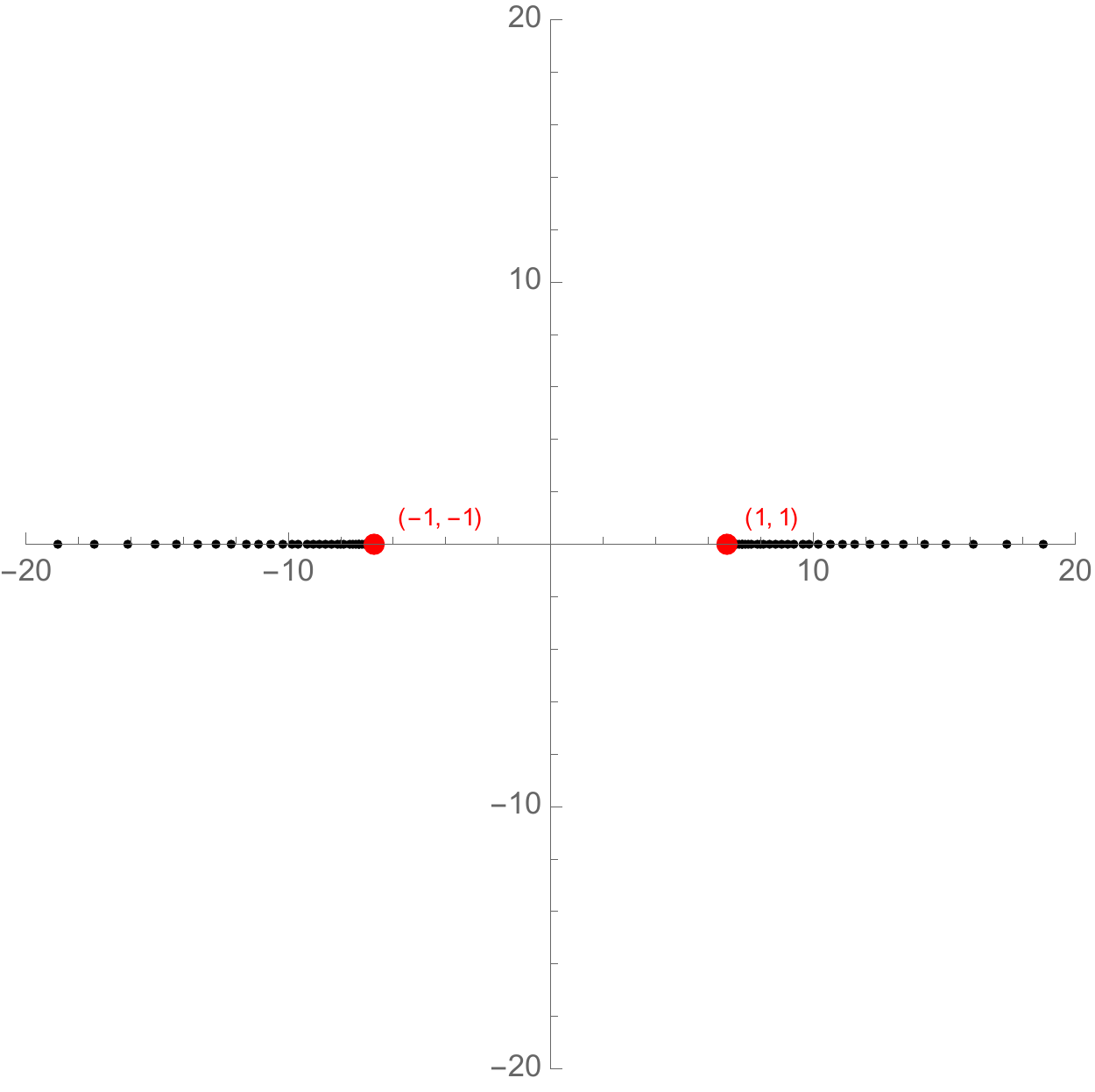}}
  \caption{Poles of the Borel--Pad\'e transforms, which would accumulate to branch
    cuts for the Borel transform of the quantum periods $\Pi_A(E,\hbar)$ (a) and
    $\Pi_B(E,\hbar)$ (b) at $u=0$ and $\Lambda=1$. The red points are
    the central charges of the BPS states which contribute to the
    branch points, and their electromagnetic charges are labelled
    nearby. See discussion in section~\ref{sc:SW-TBA}.}\label{fg:u0-bc}
\end{figure}

Let us look at some examples of the Borel plane of the quantum periods
for the modified Mathieu equation. In practice, to calculate the Borel transform, we use standard Borel--Pad\'e techniques, i.e.\ we use a finite number of terms in the 
formal power series (in this case we have used $193$ terms), and in order to extend analytically the resulting function, 
we use a Pad\'e transform of the Borel transform. In this method, branch cuts of the Borel transform 
are indicated by a dense accumulation of poles of the Borel--Pad\'e transform along a segment. The first example is when
$u=E=0$. We plot the poles of the Borel--Pad\'e transforms of
$\Pi_A(0,\hbar), \Pi_B(0,\hbar)$ in the Borel plane in
Figures~\ref{fg:u0-bc}. They indicate the existence of four branch cuts
in the case of the $A$ period, and two branch cuts in the case of
$B$ period. Since in both cases there are branch cuts along the positive
real axis, neither of the two quantum periods are Borel summable.

\begin{figure}
  \centering
  \subfloat[$\Pi_A$]{\includegraphics[width=0.4\linewidth,valign=c]
    {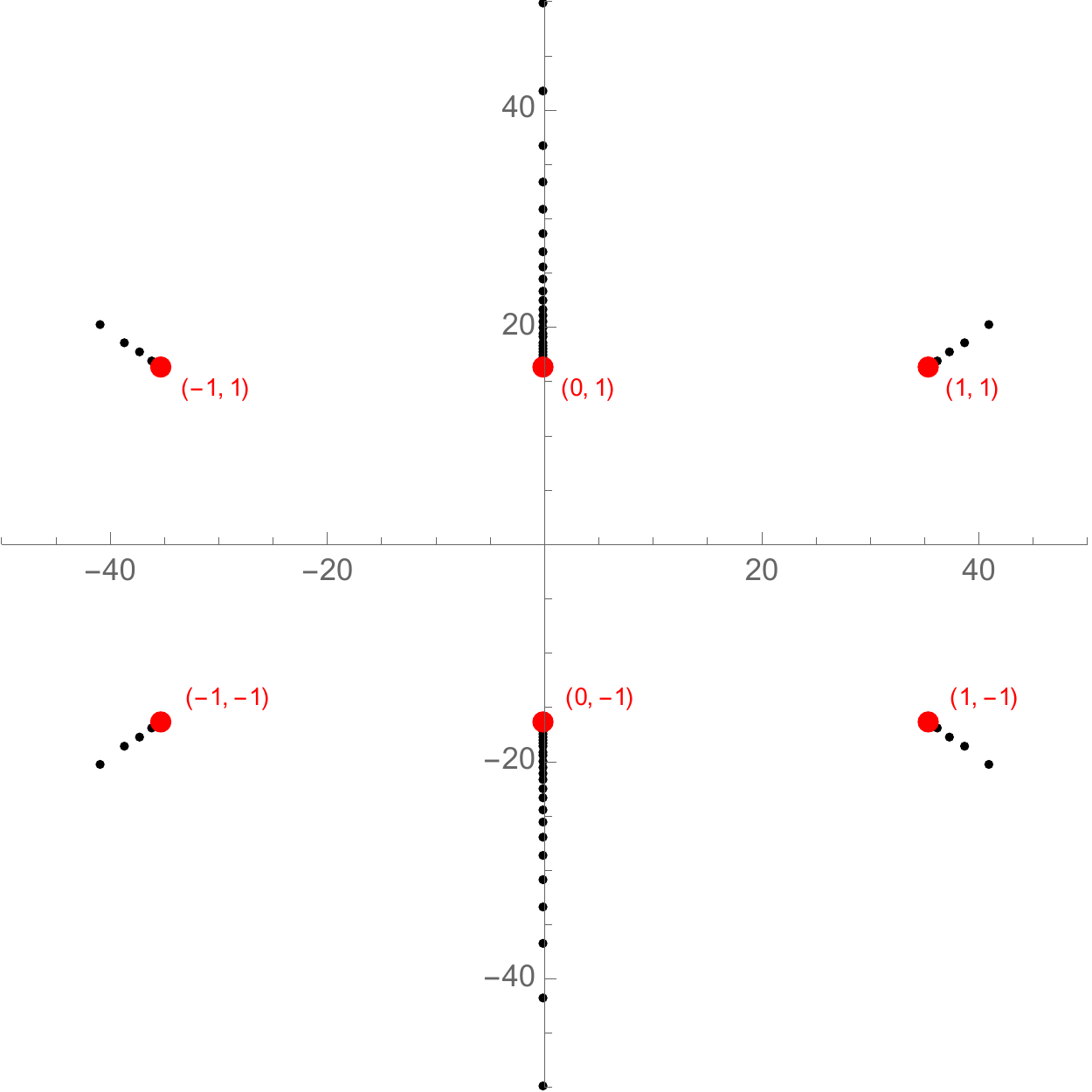}}
  \hspace{4ex}
  \subfloat[$\Pi_B$]{\includegraphics[width=0.4\linewidth,valign=c]
    {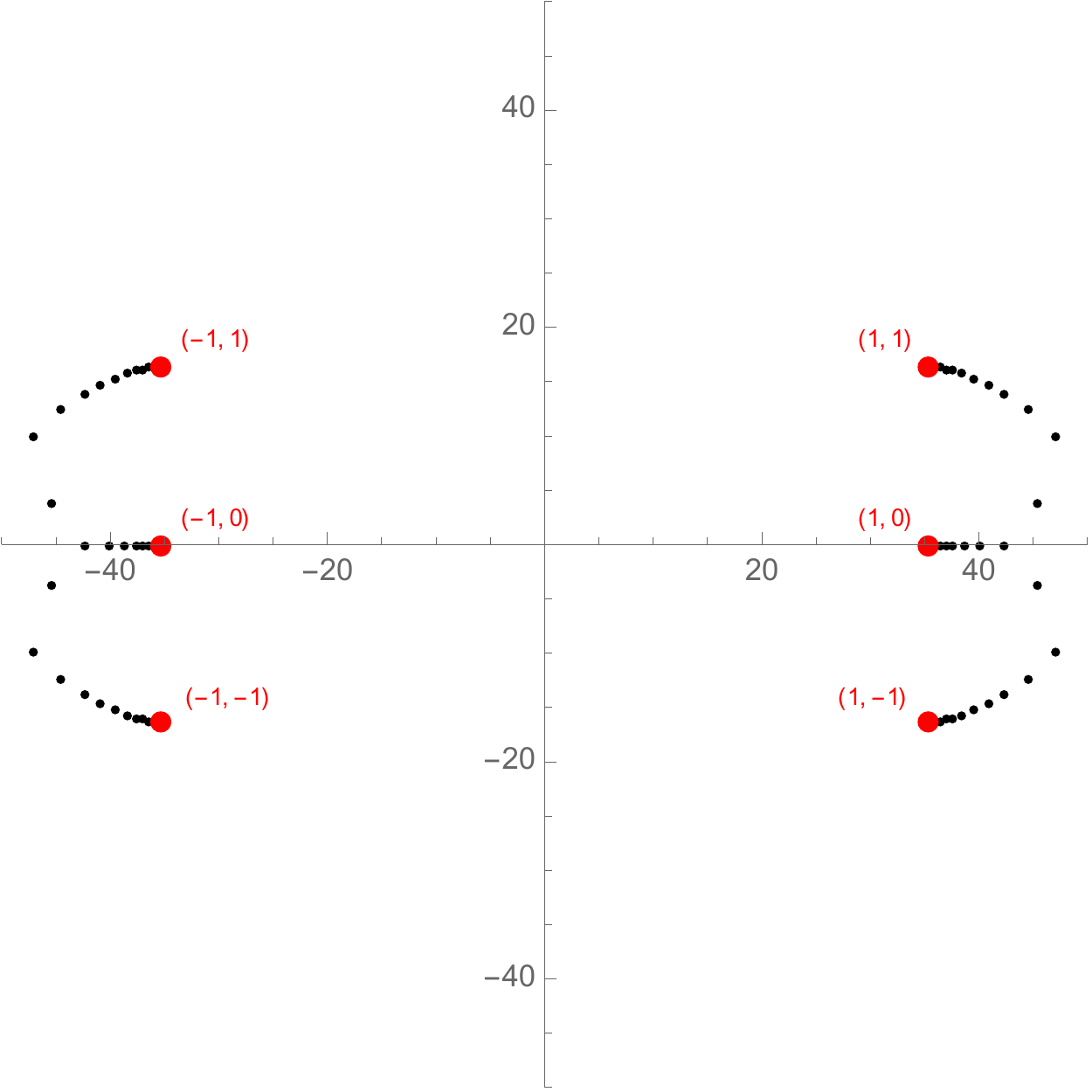}} 
  \caption{Poles of the Borel--Pad\'e transforms, which would accumulate to branch
    cuts for the Borel transforms of the quantum periods $\Pi_A(u,\hbar)$ (a) and
    $\Pi_B(u,\hbar)$ (b) at $u=4$ and $\Lambda=1$. The red points are
    the central charges of the BPS states which contribute to the
    branch points, and their electromagnetic charges are labelled
    nearby. See discussion in section~\ref{sc:SW-TBA}.}\label{fg:u4-bc}
\end{figure}

Next, we consider $u=E/2=4$. Again we plot the poles of the
Borel--Pad\'e transforms of $\Pi_A(4,\hbar), \Pi_B(4,\hbar)$ in the
Borel plane in Figures~\ref{fg:u4-bc}. In both cases we observe six
branch cuts, and they are in different locations as compared to what
we found at $u=0$. In this case, the quantum $A$ period is Borel
summable, but the quantum $B$ period is not.

As we can see, in general, the quantum periods are not Borel summable,
and their Borel transforms and resummations have a rich
structure. Fortunately the connection with SW theory gives very
powerful information on this structure, which we will explore in
detail in the next section.

\section{Quantum periods from TBA equations}\label{tbagmn}

In this section we study the TBA equations which control the analytic
properties of the quantum periods of the modified Mathieu equation. We set
$\Lambda=1$ throughout the section.

\subsection{Review of the TBA equations of Gaiotto--Moore--Neitzke}

The TBA equations we will obtain are conformal limits \cite{oper} of
the integral equations proposed by Gaiotto--Moore--Neitzke (GMN) in
\cite{gmn} to describe the hyperK\"ahler metric on the Coulomb branch
of $\CN=2$ theories compactified on $\IR^3 \times \IS^1_R$, where $R$
is the compactification radius. We will now review some basic aspects
of these equations which will be useful in the following.  The basic
ingredients in these equations are the central charges of the $\CN=2$
supersymmetric gauge theory
\begin{equation}
  Z(\bs{u})=\left(\bs{a},  \bs{a}_D \right), 
\end{equation}
where
\begin{equation}
  \bs{a}_D={\partial F_0 \over \partial \bs{a}}. 
\end{equation}
%
We define the period associated to a vector
$\bs{\gamma}\in\Gamma$ in the lattice of electromagnetic charges as
%
\begin{equation}
  Z_{\bs{\gamma}}= Z(\bs{u}) \cdot {\boldsymbol{\gamma}}.
\end{equation}
This is just a linear combination of $A$ periods and $B$ periods. 

To such a central charge we associate a ray
\begin{equation}
  \ell_{\bs{\gamma}}= \left\{ \zeta: {Z_{\bs{\gamma}}(\bs{u}) \over
      \zeta} \in \IR_-\right\}.  
\end{equation}
%
%
%
The {\it semiflat coordinate} on the Coulomb branch is given by
\begin{equation}
\chi_{\bs{\gamma}}^{\rm sf}(\zeta)= \exp\left[ \pi R \zeta^{-1}
  Z_{\bs{\gamma}} + \ri \theta_{\bs{\gamma}} + \pi R \overline \zeta
  Z_{\bs{\gamma}} \right],  
\end{equation}
where $R$ is the compactification radius, and 
\begin{equation}
 \theta_{\bs{\gamma}}= \boldsymbol{\theta} \cdot {\bs{\gamma}}
 \end{equation}
 is the angular coordinate on the fiber.  The semiflat coordinate is
 the ``uncorrected" or ``classical" coordinate, and it is corrected by
 exponentially small effects in the large $R$ limit. These
 effects are encoded in a non-linear, TBA-like integral equation,
 which reads as
\begin{equation}
\label{tba-equation}
\chi_{\bs{\gamma}}(\zeta)=\chi_{\bs{\gamma}}^{\rm sf}(\zeta)\exp\left[
  -{1\over2 \pi \ri} \sum_{\bs{\gamma}' \in \Gamma}  
\Omega(\bs{\gamma}'; \bs{u}) \langle \bs{\gamma}, \bs{\gamma}'\rangle
\CI_{\bs{\gamma}'} (\zeta) \right],  
\end{equation}
where $\Omega(\gamma;u)$ is the number of BPS states with
electromagnetic charge $\gamma$ at the point $u$ of the Coulomb branch, and
\begin{equation}
  \CI_{\bs{\gamma}}= \int_{\ell_{\bs{\gamma}}} {\rd \zeta' \over
    \zeta'} {\zeta '+\zeta \over \zeta'- \zeta} \log\left(1-
    \sigma(\bs{\gamma}) \chi_{\bs{\gamma}} (\zeta') \right).  
\end{equation}
Here, $\sigma(\bs{\gamma})$ is the quadratic refinement. It has been
argued in \cite{gmn2} that, for BPS hypermultiplets/vectormultiplets,
one has, respectively,
\begin{equation}
  \sigma(\bs{\gamma})=\mp 1. 
\end{equation}
%
%
%
%
%
We have used the normalization of \cite{cdorey}, which is more
appropriate for our normalization of charges/periods. An important
feature of \eqref{tba-equation} is that only those states whose charge
$\bs{\gamma'}$ has a non-vanishing Dirac pairing with $\bs{\gamma}$
contributes to the equation of $\chi_{\bs{\gamma}}(\zeta)$.  The
quantities $\chi_{\bs{\gamma}}(\zeta)$ characterize in a precise way
the hyperK\"ahler metric of the moduli space of the $\CN=2$ theory
compactified on $\IR^3 \times \IS^1_R$, and they can be realized as
cluster coordinates on this moduli space \cite{gmn2}. They satisfy the
property
\begin{equation}
  \label{product}
  \chi_{\bs{\gamma}+\bs{\gamma}'}(\zeta) =
  \chi_{\bs{\gamma}}(\zeta)\chi_{\bs{\gamma}'}(\zeta). 
\end{equation}

Very often we have both charges $\pm \bs{\gamma}$ appearing in the sum
in the r.h.s. of (\ref{tba-equation}). If $\theta_{\bs{\gamma}}=0$, we
have an extra symmetry \cite{oper},
\begin{equation}
\label{sym}
\chi_{\bs{\gamma}}(\zeta)=\chi_{-\bs{\gamma}}(-\zeta), 
\end{equation}
and we can combine
 \begin{equation}
   \ba
   \label{cintegral}
   \CC_{\bs{\gamma}}= \CI_{\bs{\gamma}} -\CI_{-\bs{\gamma}}
   &=
   \int_{\ell_{\bs{\gamma}}} {\rd \zeta' \over \zeta'} {\zeta '+\zeta
     \over \zeta'- \zeta} \log\left(1- \sigma(\bs{\gamma})
     \chi_{\bs{\gamma}} (\zeta') \right)-
   \int_{\ell_{-\bs{\gamma}}} {\rd \zeta' \over \zeta'} {\zeta '+\zeta
     \over \zeta'- \zeta} \log\left(1- \sigma(-\bs{\gamma})
     \chi_{-\bs{\gamma}} (\zeta') \right)\\
   &= \int_{\ell_{\bs{\gamma}}} {\rd \zeta' \over \zeta'} \left(
     {\zeta '+\zeta \over \zeta'- \zeta}-{\zeta '-\zeta \over \zeta'+
       \zeta} \right) \log\left(1- \sigma(\bs{\gamma})
     \chi_{\bs{\gamma}} (\zeta') \right)\\
   &=4 \zeta  \int_{\ell_{\bs{\gamma}}} {\rd \zeta' \over \left(
       \zeta' \right)^2-\zeta^2} \log\left(1- \sigma(\bs{\gamma})
     \chi_{\bs{\gamma}} (\zeta') \right).  
   \ea 
\end{equation}
In going from the first to the second line we have changed variables
$\zeta'\rightarrow -\zeta'$, and we used the symmetry (\ref{sym}).
 
In order to put the equations in a form similar to the TBA equations,
we will perform a change of variables akin to the one made in
\cite{gmn}. If
\begin{equation}
  Z_{\bs{\gamma}}= \re^{\ri \phi'} \left|  Z_{\bs{\gamma}} \right|
\end{equation}
then we change variables in (\ref{cintegral}) as follows:
\begin{equation}
  \zeta= -\re^{\ri \phi -\theta}, \qquad \zeta'=  -\re^{\ri \phi' -\theta'}, 
\end{equation}
and we obtain
\begin{equation}
  \CC_{\bs{\gamma}}= 2\int_\IR { \log\left(1- \sigma(\bs{\gamma})
      \chi_{\bs{\gamma}} (\theta') \right) \over \sinh(
    \theta-\theta'+ \ri \phi'-\ri \phi)} \rd \theta'.  
\end{equation}

\subsection{TBA equations for the modified Mathieu equation}
\label{sc:SW-TBA}

Building on \cite{oper, ims}, we expect to have a general
correspondence between the mathematical description of BPS states in \cite{gmn,gmn2}, and the ``resurgent"
properties of the quantum periods associated to the corresponding SW curve. As noted in \cite{oper}, this correspondence involves the conformal limit of the 
TBA equations of \cite{gmn}, which is given by
\begin{equation}
  R\to 0 \ , \quad \zeta \to 0\ ,\quad  \zeta/R\;\; \text{finite} \ .
\end{equation}
In this correspondence, the classical limit of the WKB periods
$\Pi_\gamma ^{(0)}$ corresponds to the central charge $Z_\gamma$,
while the full quantum period is obtained as the logarithm of the
Coulomb branch coordinates $\chi_\gamma(\zeta)$ (in the conformal
limit). The Borel singularities of the Borel transforms
$\widehat \Pi_\gamma$ are closely related to the BPS spectrum of the
theory, and the Stokes discontinuities of the quantum periods are
closely related to the so-called Kontsevich--Soibelman
symplectomorphisms \cite{ks, gmn,gmn2}. This correspondence is
summarized in Table \ref{table-cor}, and it can be used to obtain
integral equations of the TBA type governing the quantum periods. We
will now apply this correspondence to obtain such equations for the
modified Mathieu operator.

\begin{table}
\begin{center}
\begin{tabular}{|l | r|}
\hline
Resurgence & BPS states \\ \hline\hline
WKB curve &  SW curve\\ \hline
classical limit $\Pi_\gamma^{(0)}$ & central charge $Z(\gamma)$  \\ \hline
quantum period $\Pi_\gamma$ &  cluster coordinate $\log \, \chi_\gamma$ \\\hline
Borel singularities  & BPS spectrum  \\\hline
Stokes discontinuities  & KS symplectomorphisms  \\\hline
\end{tabular}
\end{center}
\caption{Correspondence between the mathematical structures in the
  resurgent approach to the WKB method, and those in the theory of BPS
  states.}
\label{table-cor}
\end{table}

Let us then consider the SW theory \cite{sw}, i.e.\ pure $\CN=2$ SYM with
gauge group $SU(2)$. We will denote the charge by 
\begin{equation}
  \bs{\gamma}=\gamma=(n_e, n_m). 
\end{equation}
We will use the conventions of \cite{cdorey} for the symplectic product, 
\begin{equation}
  \langle \gamma, \gamma'\rangle = \langle (n_e, n_m), (n'_e, n'_m)
  \rangle= -n_e n_m'+ n_m n'_e.  
\end{equation}
We will denote
\begin{equation}
  \chi_{e} (\zeta)= \chi_{(1,0)} (\zeta), \qquad  \chi_{m} (\zeta)=
  \chi_{(0,1)} (\zeta), \qquad \chi_{d} (\zeta)= \chi_{(1,1)} (\zeta),  
\end{equation}
and because of (\ref{product}) we have
\begin{equation}
  \chi_{d} (\zeta)= \chi_{e} (\zeta) \chi_{m} (\zeta).
\end{equation}
We will write TBA equations for $ \chi_{e} (\zeta)$ and
$ \chi_{m} (\zeta)$, as in \cite{cdorey}. We have
\begin{equation}
  \label{su2tba}
  \ba
  \chi_e (\zeta)&=  \chi^{\rm sf}_{e} (\zeta)\exp\left[-{1\over 2 \pi
      \ri} \sum_{\gamma'} c_e(\gamma') \CI_{\gamma'}(\zeta) \right],\\ 
  \chi_m (\zeta)&=  \chi^{\rm sf}_{m} (\zeta)\exp\left[-{1\over 2 \pi
      \ri} \sum_{\gamma'} c_m(\gamma') \CI_{\gamma'}(\zeta)\right], 
  \ea
\end{equation}
where
\begin{equation}
  \label{cem}
  c_e(\gamma)= \Omega(\gamma; u) \langle (1,0), \gamma\rangle, \qquad
  c_m(\gamma)= \Omega(\gamma; u) \langle (0,1), \gamma\rangle. 
\end{equation}

In order to write the integral equations, we need to know the
structure of the BPS spectrum in SW theory. It is known that there is
a curve of marginal stability ${\cal C}$ in the Coulomb branch of the
SW theory, separating a strong coupling region or chamber ${\cal S}$
inside ${\cal C}$, from a weak coupling region or chamber ${\cal W}$
outside ${\cal C}$ \cite{sw,fb,selfdual}. As we move from the strong
coupling region to the weak coupling region, the spectrum of BPS
states changes drastically by the famous wall-crossing phenomenon. We
consider the two chambers in turn.

\subsubsection{Strong coupling region}

We start with the region $\CW$ inside the curve of marginal
stability. The
spectrum consists of one monopole with charge
\begin{equation}
  \gamma_m=(0,1)
\end{equation}
and one dyon with charge
\begin{equation}
  \gamma_d=(1,1), 
\end{equation}
see \cite{sw,fb,selfdual} (we follow the conventions in \cite{fb}). We
also have the corresponding antiparticles, carrying opposite
charges. Then, the only nonzero coefficients in (\ref{su2tba}) are
\begin{equation}
  c_e(\gamma_m)=c_e(\gamma_d)=-1, \qquad c_m(\gamma_d)=1.
\end{equation}
Therefore, the equations (\ref{su2tba}) read
\begin{equation}
  \ba
  \chi_e (\zeta)&=  \chi^{\rm sf}_{e} (\zeta)\exp\left[{1\over 2 \pi
      \ri} \left( \CC_{\gamma_m}+ \CC_{\gamma_d} \right) \right],\\ 
  \chi_m (\zeta)&=  \chi^{\rm sf}_{m} (\zeta)\exp\left[-{1\over 2 \pi
      \ri} \CC_{\gamma_d} \right], 
  \ea
\end{equation}
and it is better to write them in terms of $\chi_d$, $\chi_m$,
\begin{equation}
  \ba
  \chi_d (\zeta)&=  \chi^{\rm sf}_{d} (\zeta)\exp\left[{1\over 2 \pi
      \ri}  \CC_{\gamma_m} \right],\\ 
  \chi_m (\zeta)&=  \chi^{\rm sf}_{m} (\zeta)\exp\left[-{1\over 2 \pi
      \ri} \CC_{\gamma_d} \right]. 
  \ea
\end{equation}

We now write the central charges
\begin{equation}
  Z_{d}=\re^{\ri \phi_d}|Z_d|, \qquad Z_m=-\ri \re^{\ri \phi_m} |Z_m|. 
\end{equation}
These conventions are such that, when $u \in \IR$ inside the curve of
marginal stability, we have $\phi_d=\phi_m=0$. Let us define the
functions $\epsilon_{d,m}$ and $\widetilde \epsilon_{d,m}(\theta)$ as
follows (this is similar to the notation used in \cite{amsv,ims}):
\begin{equation}
  \ba
  \chi_d\left( -\re^{\ri \phi_d -\theta}\right)&=
  \exp\left(-\epsilon_d (\theta-\ri \phi_d) \right)=
  \exp\left(-\widetilde \epsilon_d (\theta) \right),\\ 
  \chi_m\left( \ri \re^{\ri \phi_m -\theta}\right)&=
  \exp\left(-\epsilon_m (\theta-\ri \phi_m) \right)=
  \exp\left(-\widetilde \epsilon_m (\theta) \right).  
  \ea
\end{equation}
Then, the conformal limit of the TBA equations reads:
\begin{equation}\label{eq:cTBA-1}
  \ba
  \widetilde \epsilon_d(\theta)&=\pi |Z_d| \re^{\theta}-2 \int_\IR
  {\widetilde L_m(\theta') \over \cosh(\theta-\theta'+ \ri \phi_m-\ri
    \phi_d)} {\rd \theta' \over 2 \pi}, \\ 
  \widetilde \epsilon_m(\theta)&=\pi |Z_m| \re^{\theta}-2 \int_\IR
  {\widetilde L_d(\theta') \over \cosh(\theta-\theta'+ \ri \phi_d-\ri
    \phi_m)} {\rd \theta' \over 2 \pi},  
  \ea
\end{equation}
where we have shifted $\theta \to \theta -\log R$, and
\begin{equation}
  \widetilde L_{m,d}(\theta) = \log\left(1+\re^{-\widetilde
      \epsilon_{m,d}(\theta)} \right).  
\end{equation}
We have used here the fact that the BPS spectrum consists of
hypermultiplets, therefore $\sigma(\gamma)=-1$.

The equations simplify further when $u$ is real, i.e.
$u \in \IR \cap \CW=[-1, 1]$. Then one has $\phi_d=\phi_m=0$, i.e.
\begin{equation}
  Z_{\gamma_d}>0, \qquad Z_{\gamma_m}=-\ri |Z_{\gamma_m}|, 
\end{equation}
and we obtain, 
\begin{equation}\label{eq:cTBA-1R}
  \ba
  \epsilon_d(\theta)&=\pi |Z_d| \re^{\theta}-2 \int_\IR { L_m(\theta')
    \over \cosh(\theta-\theta')} {\rd \theta' \over 2 \pi}, \\ 
  \epsilon_m(\theta)&=\pi |Z_m| \re^{\theta}-2 \int_\IR { L_d(\theta')
    \over \cosh(\theta-\theta')} {\rd \theta' \over 2 \pi}.
  \ea
\end{equation}
We also note that, before taking the conformal limit, we find the more
conventional TBA equations
\begin{equation}
  \ba
  \epsilon_d(\theta)&= \pi r Z_d \cosh(\theta)-2 \int_\IR
  {L_m(\theta') \over \cosh(\theta-\theta')} {\rd \theta' \over 2
    \pi}, \\ 
  \epsilon_m(\theta)&= \pi r |Z_m| \cosh(\theta)-2 \int_\IR
  {L_d(\theta') \over \cosh(\theta-\theta')} {\rd \theta' \over 2
    \pi},  
  \ea
\end{equation}
where 
\begin{equation}
  r=2 R. 
\end{equation}
The definition of $r$ is such that we have the same conventions as in
\cite{kl-mel-1}.

The TBA equations simplify greatly when $u=0$. In this case, we have
that
\begin{equation}
  |Z_d|= |Z_m|=\xi, 
\end{equation}
and that\footnote{Anticipating the identification
  with quantum periods, this equation does not mean that the dyonic
  and magnetic quamtum periods $\Pi_D(u,\hbar), \Pi_B(u,\hbar)$ are
  identical at $u=0$, as $\hbar$ is identified with $\theta$
  differently, c.f.~\eqref{eq:TBA-dict-1}.}
\begin{equation}
  \epsilon_d(\theta)= \epsilon_m(\theta)=\epsilon(\theta). 
\end{equation}
The two TBA equations collapse to one, 
\begin{equation}
  \label{u0TBA}
  \epsilon(\theta)=\pi \xi \re^{\theta}-2 \int_\IR {L(\theta') \over
    \cosh(\theta-\theta')} {\rd \theta' \over 2 \pi}, 
\end{equation}
which coincides with the integral equation (\ref{zmathieu})
associated to the modified Mathieu equation and the Sinh-Gordon model
and studied by Zamolodchikov (the factor $\xi$ can be absorbed in a
redefinition of the angle $\theta$). The equation (\ref{u0TBA}) was
written down in \cite{oper} as governing the quantum periods at $u=0$.

We claim that the functions $\widetilde \epsilon_{d,m}(\theta)$ are
identified with quantum periods as follows
\begin{equation}\label{eq:TBA-dict-1}
  \begin{aligned}
    \widetilde \epsilon_d(x+\ri\phi_d)
    &= \frac{1}{\hbar}s(\Pi_D)(\hbar),\\
    \widetilde \epsilon_m\left(x+\ri\phi_m-\frac{\ri\pi}{2}\right)
    &= \frac{1}{\hbar} s(\Pi_B)(\hbar),
  \end{aligned}
\end{equation}
with 
\begin{equation}
  \hbar = \pi^{-1}\re^{-x}, \quad \Pi_D=\Pi_A+\Pi_B \ ,
\end{equation}
where $\Pi_D$ denotes the dyonic quantum period. Then the TBA
equations \eqref{eq:cTBA-1} are consistent with the leading order
contribution by the classical periods in the small $\hbar$ expansion
\begin{equation}
  s(\Pi_{D,B})(\hbar) = Z_{d,m} + \mc O(\hbar^2) \ .
\end{equation}
Furthermore, the TBA equations \eqref{eq:cTBA-1} clearly indicate that
for some argument angles of $\hbar$ the quantum periods have
discontinuities. These discontinuities are determined by the BPS
spectrum of SW theory and give the singularity structure of the Borel
transform of the quantum periods. These Stokes discontinuities can
also be deduced from \eqref{eq:cTBA-1}. The location of the
singularities in the Borel plane, as well as the precise
discontinuities, can be checked against the asymptotic series of the
quantum periods, by inspecting the Borel plane and by performing
lateral Borel resummations, respectively.

For instance, from the TBA equations \eqref{eq:cTBA-1}, we conclude
that $s(\Pi_B)$ are discontinuous across the rays
$\arg(\hbar) = \phi_d, \phi_d+\pi$, with 
\begin{align}
  \text{disc}_{\phi_d(+\pi)} (\Pi_B)(\hbar)
  =
  & +2\hbar\log\left(1+\re^{-\tfrac{1}{\hbar} s(\Pi_D)(\hbar)}\right) \nn
    =
  & 2\hbar\(\re^{-\tfrac{1}{\hbar} s(\Pi_D)(\hbar)} -
    \frac{1}{2}\re^{-2\tfrac{1}{\hbar} s(\Pi_D)(\hbar)} + \ldots\) \ .
    \label{eq:disc-B1}
\end{align}
When $u=0$, we have $\phi_d=0$, and the discontinuities are located at
$\arg(\hbar) = 0,\pi$. 
The discontinuity across the ray $\arg(\hbar) = 0$ can be computed by
a lateral Borel resummation of the quantum $B$ period. We check it
against the right hand side of \eqref{eq:disc-B1}, and find good
agreement. See Table~\ref{tb:disc-Bu0}. Similarly, $s(\Pi_D)$ is
discontinuous across the rays $\arg(\hbar) = \phi_m\pm\frac{\pi}{2}$,
and one has
\begin{equation}
  \text{disc}_{\phi_m\pm\frac{\pi}{2}} (\Pi_D)(\hbar)
  = -2\hbar
  \log\left(1+\re^{-\tfrac{1}{\hbar} s(\Pi_B)(\hbar)}\right) \ .
\end{equation}
 Numerical checks for this discontinuity formula are
completely analogous.

\begin{table}
  \centering
  \begin{tabular}{cll}\toprule
    terms & lateral Borel sum & r.h.s.\ of \eqref{eq:disc-B1}\\\midrule
    181   & 0.17499253901611 & 0.1749925390148815032360 \\
    185   & 0.17499253901578 & 0.1749925390148815032482 \\
    189   & 0.17499253901545 & 0.1749925390148815032553 \\
    193   & \underline{0.174992539015}19
                              & \underline{0.17499253901488150325}95 \\\bottomrule
  \end{tabular}
  \caption{Discontinuity across the ray $\arg(\hbar) = 0$
    for $\Pi_B(u=0,\hbar=1)$ computed by lateral Borel resummation and
    by using \eqref{eq:disc-B1} with increasing number of terms in the
    asymptotic series. Underlined are stabilised digits.}
  \label{tb:disc-Bu0}
\end{table}

From the discontinuity formula \eqref{eq:disc-B1} we can deduce in the
standard way a formula for the large order behavior of $\Pi_B^{(n)}$,
of the form (see e.g. \cite{msw})
\begin{align}
&  \ri \Pi_B^{(n)}  \sim
  \frac{2 A^{-2n+b}}{\pi} \Gamma(2n+b) \nn
  &\cdot\(
    1-\frac{\mu_2 A}{2n+b-1}+ \frac{\mu_3
    A^2}{(2n+b-1)(2n+b-2)} - \frac{\mu_4
    A^3}{(2n+b-1)(2n+b-2)(2n+b-3)}\cdots\). \label{eq:aD-asymp}
\end{align}
If we write the dyonic quantum period as
\begin{equation}
  \Pi_D= \sum_{n \ge 0} \Pi_D^{(n)}\hbar^{2n},
\end{equation}
we can identify
\begin{equation}
  A= \Pi_D^{(0)}, \quad b=-1,\quad \mu_2= \Pi_D^{(1)}, \quad \mu_3 =
  (\Pi_D^{(1)})^2/2 
  ,\quad \mu_4 = (\Pi_D^{(1)})^3/6+\Pi_D^{(2)}  ,\quad\ldots
\end{equation}
These identities are numerically checked at $u=0$ up to
all stabilised digits (more than 40) with the help of Richardson
transforms.

The large order behavior of $\Pi_B^{(n)}$ also indicates that the
Borel transform $\widehat \Pi_B(\zeta)$ has branch points at
$\zeta = \pm \Pi_D^{(0)}$, which is the central charge of the BPS
state (dyon) whose electromagnetic charge has non-vanishing Dirac
pairing with the charge of the monopole. Similarly, the Borel
transform of $\Pi_A(u,\hbar)$ should have branch points at the central
charges of monopoles and dyons with electromagnetic charges
$\pm(0,1), \pm(1,1)$, while the Borel transform of $\Pi_B(u,\hbar)$
have branch points only at the central charges of dyons. This explains
the Borel plane plots in Figure~\ref{fg:u0-bc}, where we also
superimpose the central charges of the contributing BPS states as red
spots.

\subsubsection{Weak coupling region}

Let us now consider the region outside the curve of marginal
stability. The spectrum consists of dyons with charge $\pm \gamma_n$,
where
\begin{equation}
  \gamma_n= (n,1), \qquad n \in \IZ, 
\end{equation}
and $W$ boson with charges $\pm \gamma_e$, where
\begin{equation}
  \gamma_e=(1,0). 
\end{equation}
From (\ref{cem}) we conclude that
\begin{equation}
  c_e(\gamma_n)=-1
\end{equation}
and
\begin{equation}
  c_m(\gamma_n)=n, \qquad c_m(\gamma_e)=2, 
\end{equation}
where we used the fact that 
\begin{equation}
\Omega(\gamma_e;u)=2
\end{equation}
in the weak coupling region.

As in \cite{cdorey}, we write the equations for $\chi_e$, $\chi_m$. We
find
\begin{equation}
  \ba
  \chi_m(\zeta)&= \chi_m^{\rm sf}(\zeta) \exp\left[ -{1\over  \pi \ri}
    \CC_{\gamma_e}(\zeta) -{1\over 2 \pi \ri} \sum_{n \in \IZ} n
    \CC_{\gamma_n}(\zeta) \right],\\ 
  \chi_e(\zeta)&=\chi_e^{\rm sf}(\zeta) \exp\left[ {1\over 2 \pi \ri}
    \sum_{n \in \IZ} \CC_{\gamma_n}(\zeta) \right].  
  \ea
\end{equation}
In this region we will write
\begin{equation}
  Z_m= \ri \re^{\ri \phi_m} |Z_m|, \qquad Z_e= \re^{\ri \phi_e} |Z_e|,
  \qquad Z_{\ell}=\re^{\ri \phi_\ell} |Z_\ell|,  
\end{equation}
where we have denoted $Z_\ell= Z_{\gamma_\ell}$ the central charge of
a dyon. This is chosen in such a way that, if $u$ is real, we have
$\phi_e=\phi_m=0$.  We now define %
\begin{equation}
  \ba
  \chi_e\left( -\re^{\ri \phi_e -\theta}\right)&=
  \exp\left(-\epsilon_e (\theta-\ri \phi_e) \right)=
  \exp\left(-\widetilde \epsilon_e (\theta) \right),\\ 
  \chi_m\left( -\ri \, \re^{\ri \phi_m -\theta}\right)&=
  \exp\left(-\epsilon_m (\theta-\ri \phi_m) \right)=
  \exp\left(-\widetilde \epsilon_m (\theta) \right), \\ 
  \chi_\ell \left( - \, \re^{\ri \phi_\ell -\theta}\right)&=
  \exp\left(-\epsilon_\ell (\theta-\ri \phi_\ell) \right)=
  \exp\left(-\widetilde \epsilon_\ell (\theta) \right),   
  \ea
\end{equation}
We then obtain the equations,
\begin{equation}\label{eq:cTBA-2}
  \ba
  \widetilde \epsilon_e(\theta)
  &=\pi |Z_e|  \re^\theta+{1\over \pi} \int_\IR {\widetilde L_m
    (\theta') \over \cosh(\theta-\theta'+\ri \phi_m- \ri \phi_e)} \rd
  \theta' 
  + 
  {1\over \pi \ri}  \sum_{\ell \not=0} \int_\IR {\widetilde
    L_\ell(\theta') \rd \theta' \over \sinh(\theta-\theta'+\ri
    \phi_\ell- \ri \phi_e)} , \\ 
  \widetilde \epsilon_m(\theta)&=\pi |Z_m|  \re^\theta-{2\over \pi}
  \int_\IR {\widetilde L_e (\theta')\rd \theta' \over
    \cosh(\theta-\theta'+\ri \phi_e- \ri \phi_m)}-{1\over \pi}
  \sum_{\ell \in \IZ} \ell \int_\IR {\widetilde L_\ell(\theta') \rd
    \theta' \over  \cosh(\theta-\theta'+\ri \phi_\ell- \ri \phi_m)} 
  \ea
\end{equation}
where
\begin{equation}
  \widetilde L_{e,m}(\theta) = \log\left(1\mp \re^{-\widetilde
      \epsilon_{e,m}(\theta)} \right), \qquad \widetilde
  L_{\ell}(\theta) = \log\left(1+ \re^{-\widetilde
      \epsilon_{\ell}(\theta)} \right). 
\end{equation}
Here we have assumed that
\begin{equation}
  \sigma(\gamma_e)=1, 
\end{equation}
since the $W$ boson is a vector multiplet \cite{gmn2}. In the equation
for $\widetilde \epsilon_e(\theta)$ we have written down explicitly
the term corresponding to the dyon with zero electric charge
$\gamma_{\ell=0}=\gamma_m$, which is the magnetic monopole.  We can
also deduce the TBA equation for $\widetilde \epsilon_{\ell}(\theta)$,
by combining the two equations above. We find
\begin{equation}
\label{dyon-eps}
\widetilde \epsilon_\ell (\theta)=\pi |Z_\ell|  \re^\theta+{2\over \pi
  \ri} \int_\IR {\widetilde L_e (\theta')  \rd \theta' \over
  \sinh(\theta-\theta'+\ri \phi_e- \ri \phi_\ell)} +  
{1\over \pi \ri}  \sum_{k\in \IZ} (\ell-k) \int_\IR {\widetilde
  L_k(\theta') \rd \theta' \over \sinh(\theta-\theta'+\ri \phi_k- \ri
  \phi_\ell)}.  
\end{equation}
It is useful to isolate the contribution from the magnetic monopole
$k=0$ explicitly in the last term, so that we obtain
\begin{equation}
  \ba
  \label{dyon-eps2}
  \widetilde \epsilon_\ell (\theta)&=\pi |Z_\ell|  \re^\theta+{2\over
    \pi \ri} \int_\IR {\widetilde L_e (\theta')  \rd \theta' \over
    \sinh(\theta-\theta'+\ri \phi_e- \ri \phi_\ell)}  +  
  {1\over \pi \ri}  \sum_{k\not=0} (\ell-k) \int_\IR {\widetilde
    L_k(\theta')  \rd \theta'\over \sinh(\theta-\theta'+\ri \phi_k-
    \ri \phi_\ell)}\\ 
  &-{\ell \over \pi}  \int_\IR {\widetilde L_m(\theta')  \rd
    \theta'\over \cosh(\theta-\theta'+\ri \phi_m- \ri \phi_\ell)}. 
  \ea
\end{equation}
The above equations have some interesting reality properties along the
real axis, where $\phi_e=\phi_m=0$. In that case, since
\begin{equation}
  Z_\ell= \ell |Z_e| + \ri |Z_m|, 
\end{equation}
one has that
\begin{equation}
  \phi_{\ell}+ \phi_{-\ell}= \pi, \qquad \ell\in \IZ. 
\end{equation}
It is then easy to see that the conjugation property
\begin{equation}
  \widetilde \epsilon_{-\ell} (\theta)= \widetilde \epsilon^*_{\ell} (\theta)
\end{equation}
is compatible with the TBA system. In addition,
$\epsilon_{e,m}(\theta)$ are real in this case.

In the weak coupling region, we propose the following
identification with quantum periods
\begin{equation}\label{eq:TBA-dict-2}
  \begin{aligned}
    \widetilde \epsilon_m\left(x+\ri\phi_m+\frac{\ri\pi}{2}\right) =
    &\frac{1}{\hbar}s(\Pi_B)(\hbar),\\
    \widetilde \epsilon_e(x+\ri\phi_e) =
    &\frac{1}{\hbar}s(\Pi_A)(\hbar),\\
    \widetilde \epsilon_\ell(x+\ri\phi_\ell) =
    &\frac{1}{\hbar}s(\Pi_\ell)(\hbar),
  \end{aligned}
\end{equation}
with 
\begin{equation}
  \hbar = \pi^{-1}\re^{-x} , \quad \Pi_{\ell}=\ell \Pi_A+ \Pi_B \ .
\end{equation}
The TBA equations \eqref{eq:cTBA-2} then imply that the Borel
transforms of $\Pi_A(\hbar), \Pi_B(\hbar)$ have branch points at the
central charges of the BPS states whose electromagnetic charges have
non-vanishing Dirac pairing with those of the W-boson and monopole,
respectively. For $\Pi_A(\hbar)$, these are the BPS states with
charges $\pm(\ell,\pm 1), \ell=0,1,2,\ldots$; for $\Pi_B(\hbar)$,
these are the BPS states with charges
$\pm(1,0), \pm(\ell,\pm 1), \ell=1,2,\ldots$. This explains the Borel
plane plots in Figure~\ref{fg:u4-bc} with $u=4$, well in the weak
coupling region. We also superimpose in the plots the central charges
of the contributing BPS states as red spots.

\begin{figure}
  \centering
  \subfloat[$\Pi_A$]{\includegraphics[width=0.4\linewidth,valign=c]
    {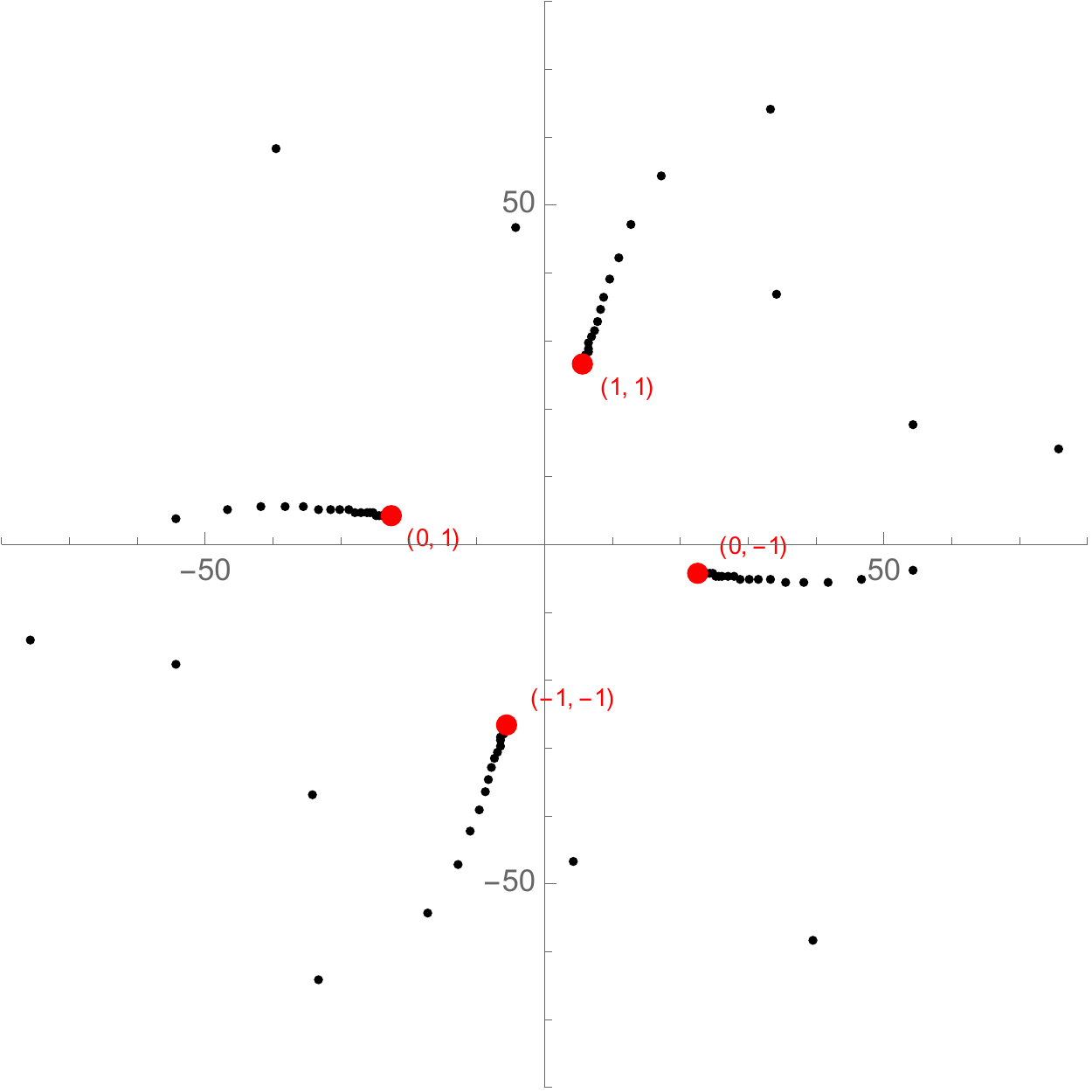}} 
  \hspace{4ex}
  \subfloat[$\Pi_B$]{\includegraphics[width=0.4\linewidth,valign=c]
    {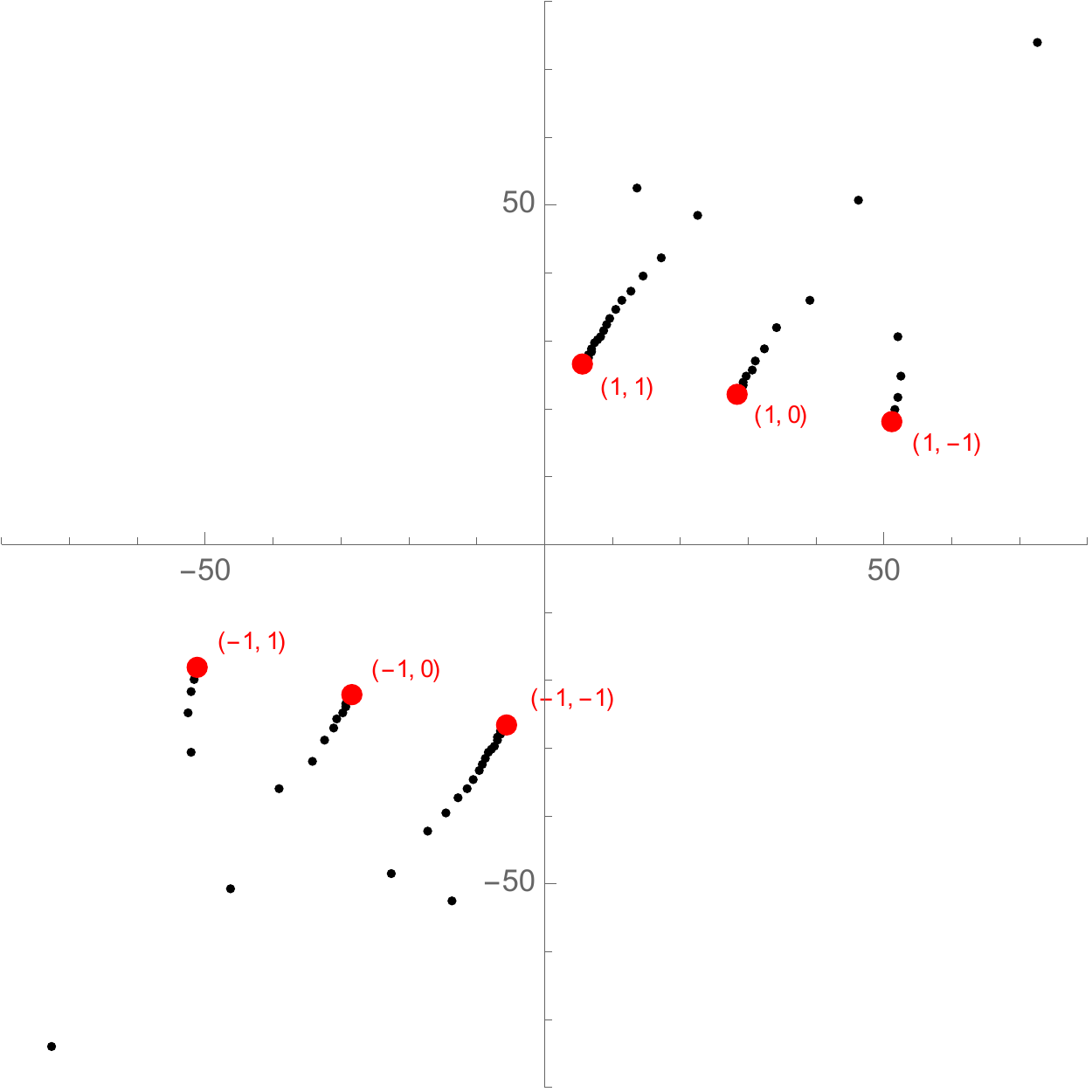}}
  \caption{Poles of the Borel--Pad\'e transforms, which would accumulate to branch
    cuts for the Borel transform of the quantum periods $\Pi_A(E,\hbar)$ (a) and
    $\Pi_B(E,\hbar)$ (b) at $u=1+4\ri$ and $\Lambda=1$. The red points
    are the central charges of the BPS states which contribute to the
    branch points, and their electromagnetic charges are labelled
    nearby.}\label{fg:ucpx-bc}
\end{figure}

In addition, the TBA equations \eqref{eq:cTBA-2} also indicate the
following discontinuities for the resummed quantum periods $s(\Pi_A)(\hbar)$,
$s(\Pi_B)(\hbar)$ in the $\hbar$-plane. The resummed quantum $A$ period
$s(\Pi_A)$ is discontinuous
\begin{itemize}
\item across the rays
$\arg(\hbar) = \phi_m\pm\frac{\pi}{2}$ with the discontinuity
\begin{equation}\label{eq:disc-A2m}
  \text{disc}_{\phi_m\pm\frac{\pi}{2}}(\Pi_A)
  = -2\hbar \log(1+\re^{-\tfrac{1}{\hbar}s(\Pi_B)(\hbar)}) \ ;
\end{equation}
\item across the rays $\arg(\hbar) = \phi_\ell(+\pi)$
with the discontinuity
\begin{equation}\label{eq:disc-A2l}
  \text{disc}_{\phi_\ell(+\pi)}(\Pi_A)
  = 2\hbar \log(1+\re^{-\tfrac{1}{\hbar}s(\Pi_\ell)(\hbar)}) \ ,\quad
  \ell\in\IZ\ .
\end{equation}
\end{itemize}
On the other hand, the quantum $B$ period $s(\Pi_B)$ is discontinuous
\begin{itemize}
\item across the rays $\arg(\hbar) = \phi_e(+\pi)$ with the discontinuity
\begin{equation}\label{eq:disc-B2e}
  \text{disc}_{\phi_e(+\pi)}(\Pi_B)
  = -4\hbar \log(1-\re^{-\tfrac{1}{\hbar}s(\Pi_A)(\hbar)}) \ ;
\end{equation}
\item across the rays $\arg(\hbar) = \phi_\ell(+\pi)$ ($\ell\neq 0$)
  with the discontinuity
\begin{equation}\label{eq:disc-B2l}
  \text{disc}_{\phi_\ell(+\pi)}(\Pi_B)
  = -2\ell\hbar \log(1+\re^{-\tfrac{1}{\hbar}s(\Pi_\ell)(\hbar)}) \ ,\quad
  \ell\neq 0\ .
\end{equation}
\end{itemize}
To test these formulae, we consider the case of $u=1+4\ri$, where the
branch cuts of the Borel transform of quantum $A$ and $B$ periods are well
separated, as seen in Figure~\ref{fg:ucpx-bc}. We compute the discontinuity
via lateral Borel resummation for various rays and
find good agreement with the r.h.s.\ of the formulae
\eqref{eq:disc-A2m}--\eqref{eq:disc-B2l}, see
Tables~\ref{tb:disc-Aucpx},\ref{tb:disc-Bucpx}.

\begin{table}
  \centering
  \resizebox{\linewidth}{!}{
  \begin{tabular}{c*{2}{>{$}l<{$}}}\toprule
    BPS state & (0,-1) & (1,1)\\\midrule
    lateral Borel sum
    & 1.77819420225\times 10^{-10}-\ri 4.0146843089\times 10^{-11}
    & 5.37838410\times 10^{-13}+\ri 3.166367886\times 10^{-12}\\\midrule
    r.h.s.\ of \eqref{eq:disc-A2m},\eqref{eq:disc-A2l}
    & 1.7781942022540\times 10^{-10}-\ri 4.014684308966\times 10^{-11}
    & 5.3783841037\times 10^{-13}+\ri 3.16636788673\times
      10^{-12}
    \\\bottomrule
  \end{tabular}}
\caption{Discontinuity of $\Pi_A(u=1+4\ri,\hbar=1)$ across the rays associated to BPS states $\gamma = (0,-1), (1,1)$, computed by
  lateral Borel resummation and by using
  \eqref{eq:disc-A2m},\eqref{eq:disc-A2l} with up to 193
  terms in the asymptotic series. Only stabilised digits are listed.}
  \label{tb:disc-Aucpx}
\end{table}

\begin{table}
  \centering
  \resizebox{\linewidth}{!}{
  \begin{tabular}{c*{2}{>{$}l<{$}}}\toprule
    BPS state & (1,1) & (1,0)\\\midrule
    lateral Borel sum
    & 5.37838\times 10^{-13}+\ri 3.166367\times 10^{-12}
    & 5.979\times 10^{-16}+\ri 4.701\times 10^{-16}\\\midrule
    r.h.s.\ of \eqref{eq:disc-B2e},\eqref{eq:disc-B2l}
    & 5.3783841\times 10^{-13}+\ri 3.16636788\times 10^{-12}
    & 5.97982\times 10^{-16}+\ri 4.70119\times
      10^{-16}
    \\\bottomrule
  \end{tabular}}
\caption{Discontinuity of $\Pi_B(u=1+4\ri,\hbar=1)$ across the rays associated to BPS states $\gamma = (1,1), (1,0)$, computed by
  lateral Borel resummation and by using
  \eqref{eq:disc-B2e},\eqref{eq:disc-B2l} with up to 193
  terms in the asymptotic series. Only stabilised digits are listed.}
  \label{tb:disc-Bucpx}
\end{table}

Finally, we would like to mention that different TBA-like equations
for the quantum periods of the modified Mathieu operator have been
proposed in \cite{ito-shu-unp} and more recently in \cite{h-neitzke}.

\subsection{Solving the TBA equations in the strong coupling
  region}\label{tbau0}

As we have argued, Borel sums of quantum periods are solutions to the TBA
equations \eqref{eq:cTBA-1},\eqref{eq:cTBA-2}. In principle the
resummed quantum periods can be computed from these TBA equations
by using the dictionaries \eqref{eq:TBA-dict-1} and
\eqref{eq:TBA-dict-2}. In practice, however, these equations are
difficult to use. First of all, one needs information on the boundary
conditions at strong coupling in order to solve the equations. In
addition, the standard tools to solve these equations numerically
converge very slowly.

Let us first consider the simplest example at $u=0$, where the TBA
system collapses to a single equation \eqref{u0TBA}, which we
reproduce here (we have absorbed the factor $\xi$ in \eqref{u0TBA} in
the angle $\theta$)
\begin{equation}\label{eq:cTBA-0}
  \epsilon(\theta) = \pi\re^{\theta} -
  \int_{\IR}\frac{\log(1+\re^{-\epsilon(\theta')})}
  {\pi\cosh(\theta-\theta')} 
  \rd \theta' \, .
\end{equation}
The solution can be identified with the quantum dyon period through
the dictionary
\begin{equation}
  \epsilon(\theta) = \frac{1}{\hbar}s(\Pi_D)(u=0,\hbar) \ ,\quad \hbar =
  \Pi^{(0)}_D \pi^{-1} \re^{-\theta}={16 \sqrt{\pi} \over \Gamma
    \left({1\over 4}\right)^2}\re^{-\theta} \ .
\end{equation}
Furthermore, the asymptotic behavior of the solution
$\epsilon(\theta)$ as $\theta \rightarrow \infty$ is of the form
\begin{equation}
  \epsilon(\theta)\sim  \pi \re^{\theta} + \sum_{n \ge 1} \epsilon^{(n)} \re^{(1-2n)\theta}, 
\end{equation}
whose coefficients are identified with the quantum corrections to the
dyon period
\begin{equation}
  \epsilon^{(n)} = \frac{1}{\pi^{2n-1}}(\Pi^{(0)}_D)^{2n-1} \Pi_D^{(n)}
  \ .
\end{equation}
 
It turns out that the equation \eqref{eq:cTBA-0} admits many possible
boundary conditions at $\theta \rightarrow -\infty$. This is in stark
contrast to the TBA equations for polynomial potentials studied in
\cite{oper, ims}, where the equations themselves fix the behavior of
the solutions at $\theta \rightarrow -\infty$. One possibility for the
boundary conditions at $\theta \rightarrow -\infty$ is the linear
behavior \eqref{zamobc}. This type of behavior was considered by
Zamolodchikov in \cite{post-zamo} but in a slightly different context,
as we will discuss in Sec. \ref{tssst} (see also
Appendix~\ref{appendixzamo}).  However, it can be seen that this is
not well suited for the quantum periods we are studying\footnote{In
  \cite{oper} it was also pointed out that \eqref{eq:cTBA-0} admits
  many boundary conditions at $\theta \rightarrow -\infty$.  However,
  it is claimed there that the correct boundary condition for the
  quantum period is precisely of the type \eqref{bcP}, namely,
  $\epsilon(\theta)\sim \theta/2+\ldots$ for $\theta\to-\infty$, which
  is not quite correct for the reasons explained here.}. One quick way
to see this is that the linear boundary condition with $P \not=0$
implies $\epsilon^{(1)}<-1/3$ (c.f. \eqref{fqc}), while from the
quantum Matone relation \eqref{eq:qMatone} we find
\begin{equation}
\label{matone-u0}
  \epsilon^{(1)} = \frac{1}{\pi}\Pi_D^{(0)}\Pi_D^{(1)} =
  \frac{1}{2\pi\ri}\(\Pi_A^{(0)}\Pi_B^{(1)} - \Pi_B^{(0)}\Pi_A^{(1)}\)
  = -\frac{1}{3}.
\end{equation}
It turns out that the appropriate boundary condition in this case is
given by
\begin{equation}\label{eq:bdy-cor}
  \epsilon(\theta)=-2 \log\left( -{2 \theta \over \pi} \right)
  +\cdots, \qquad \theta \to -\infty \ .
\end{equation}
This boundary condition for the TBA equation \eqref{eq:cTBA-0} was
also studied by Zamolodchikov in
\cite{zamo-reso}\footnote{Alternatively we can justify this boundary
  condition by using the results in section~\ref{tssst}, see equation
  \eqref{eps0}.}.  One can use a small modification of the
``dilogarithm trick" of \cite{zamo-TBA} to show that, with the
boundary condition (\ref{eq:bdy-cor}), one has indeed
(\ref{matone-u0}) (in the context of \cite{zamo-reso}, this
calculation gives the central charge $c=1$ for the corresponding
sinh-Gordon theory).

To implement numerically the boundary condition (\ref{eq:bdy-cor}), we
borrow a trick from \cite{post-zamo}. We define a continuous function
\begin{equation}
  f_1(\theta) = -2\log\left(1+\frac{2}{\pi}\log(1+\re^{-\theta})\right),
\end{equation}
 which has the same boundary behavior as
\eqref{eq:bdy-cor} and is exponentially suppressed when
$\theta \to +\infty$. We then look for a function $F_1(\theta)$ which
satisfies
\begin{equation}
  f_1(\theta) = \int_{\IR} K(\theta-\theta') F_1(\theta')\rd \theta' \
  ,\quad K(\theta) = \frac{1}{\pi\cosh(\theta)}\ .
\end{equation}
The generic solution to this linear integral equation  is
\begin{equation}
  F_1(\theta) = \frac{1}{2}\left(f_1\left(\theta+\ri\frac{\pi}{2}\right) +
  f_1 \left(\theta-\ri\frac{\pi}{2}\right)\right) \ .
\end{equation}
For our particular $f_1(\theta)$, we thus have
\begin{equation}
  F_1(\theta) =
  -\log\left(1+\frac{2}{\pi}\log(1+\ri\re^{-\theta})\right)
  -\log\left(1+\frac{2}{\pi}\log(1-\ri\re^{-\theta})\right)
  \ .
\end{equation}
This is a real function for $\theta\in\IR$. The TBA equation
\eqref{eq:cTBA-0} can then be written as
\begin{equation}\label{eq:TBA-SW2}
  \epsilon(\theta) = \pi \re^{\theta} + f_1(\theta) -
  \int_{\IR}\rd \theta'
  \frac{1}{\pi\cosh(\theta-\theta')}
  \(\log(1+\re^{-\epsilon(\theta')})+F_1(\theta')\) 
  \ ,
\end{equation}
where both boundary conditions at $\pm \infty$ are explicitly
spelt out.

The numerical solution to the
TBA equation \eqref{eq:cTBA-0} converges rather slowly, and we managed
to obtain 6 stabilised digits for $\hbar = 1$ and 7 stabilised digits
for $\hbar = 1/2$. These results, on the other hand, do agree with the
Borel resummation of the quantum dyon period. See Table~\ref{tb:TBA}.

\begin{table}
  \centering
  \begin{tabular}{cll}\toprule
    &$\hbar = 1$ & $\hbar = 1/2 $\\\midrule
    TBA & 6.62781 & 13.47880 \\    
    Borel sum & 6.62781917$\ldots$ & 13.47880936$\ldots$\\
    \bottomrule
  \end{tabular}
  \caption{Quantum dyon period at $u=0$.}\label{tb:TBA}
\end{table}


Let us now move away from the point $u=0$ but remain in
the strong coupling region with $u\in (-1,0)\cup(0,1)$. The TBA system
\eqref{eq:cTBA-1R} has two integral equations coupled to each
other. Nevertheless, at $\theta \to -\infty$ the first terms
$\propto \re^{\theta}$ on the r.h.s.\ of both equations in
\eqref{eq:cTBA-1R} are negligible, and the TBA system also collapses
to the single equation \eqref{eq:cTBA-0} (with the first term on the
r.h.s.\ suppressed). Therefore both
$\epsilon_d(\theta),\epsilon_m(\theta)$ should have the same boundary
condition as \eqref{eq:bdy-cor}, in other words
\begin{equation}
  \epsilon_d(\theta) \sim \epsilon_m(\theta) \sim
  -2\log\(-\frac{2\theta}{\pi}\) 
  +\ldots\ ,\quad \theta \to -\infty \ . 
\end{equation}
This is corroborated by the fact that the Matone relation
\eqref{matone-u0} can be reproduced with this boundary behavior by
using again a slight modification of the ``dilogarithm trick''
\cite{zamo-TBA}. We use again the trick of inserting the pair of
$f_1(\theta), F_1(\theta')$ functions, and we find that the numerical
solution to the TBA system \eqref{eq:cTBA-1R} has roughly the same
speed of convergence as the solution to \eqref{eq:cTBA-0} for
$u=0$. We tabulate the results for $u=1/3$ in
Tables~\ref{tb:dynu1o3},\ref{tb:monu1o3} and they also agree with the
Borel sum of the quantum periods. Note that the TBA system is solved
with $\theta \in \IR$, which in light of \eqref{eq:TBA-dict-1}
corresponds to real $\hbar$ for the quantum dyon period and to
imaginary $\hbar$ for the quantum monopole period.

\begin{table}
  \centering
  \begin{tabular}{c *{2}{>{$}l<{$}}}\toprule
    & \hbar = 1 & \hbar = 1/2 \\\midrule
    TBA & 9.16476 & 18.61486 \\
    Borel sum & 9.16476545\ldots & 18.61486738\ldots \\\bottomrule
  \end{tabular}
  \caption{Quantum dyon period at $u=1/3$.}
  \label{tb:dynu1o3}
\end{table}

\begin{table}
  \centering
  \begin{tabular}{c *{2}{>{$}l<{$}}}\toprule
    & \hbar = -\ri & \hbar = -\ri/2 \\\midrule
    TBA & 4.26480 & 8.716486 \\
    Borel sum & 4.26480153\ldots & 8.716486917\ldots \\\bottomrule
  \end{tabular}
  \caption{Quantum monopole period at $u=1/3$.}
  \label{tb:monu1o3}
\end{table}

\section{Quantum periods from instanton calculus}\label{res}

Instanton calculus \cite{n,ns} leads to a resummation of the quantum
periods of the modified Mathieu equation \eqref{QP}, as pointed out in
\cite{mirmor}. This produces exact functions of $\hbar$ which we will
denote by
\begin{equation}
  \Pi_{A,B}^{\rm ex}  (E,\hbar).
\end{equation}
In this section we explain this resummation in detail and we compare
it to the Borel resummation obtained in the context of the exact WKB
method.

\subsection{Review of instanton calculus}\label{ic}

Let us first review some basic ingredients of instanton calculus in
the 4d $N=2$ SYM with gauge group $G=SU(N)$ \cite{n,no2,ns,
  Flume:2002az, bfmt}.

We denote a partition (or  Young tableaux) by
\be Y=(y_1, y_2, \cdots),\ee 
 its transposed by
\be Y^t=(y_1^t, y_2^t, \cdots),\ee
and a vector of Young tableaux as%
\begin{equation}
  \label{Y-vec}
  \boldsymbol{Y}=(Y_1, \cdots, Y_N). 
\end{equation}
It is useful to  define
\begin{equation}
  h_Y(s)=y_i-j, \qquad v_Y(s)= y^t_j-i, 
\end{equation}
where $ s=(i,j) $ is a  box (not necessarily in the
partition $Y$). We will also use 
\begin{equation}\label{ly}
  \ell( \boldsymbol{Y})= \sum_{I=1}^N \ell(Y_I) \ ,
\end{equation}
whith
\begin{equation}
  \ell(Y)= \sum_i y_i \ .
\end{equation}
The four dimensional   $SU(N)$ Nekrasov partition function is \cite{n, Flume:2002az}
\be
\label{z4d}
Z (\bs{a}; \epsilon_1, \epsilon_2)=\sum_{\boldsymbol{Y}}  \left((-1)^N
  \Lambda^{2N} \right)^{\ell (\boldsymbol{Y})}  \CZ_{\boldsymbol{Y}},  
\ee
where 
\be
\ba
\label{zdef}\CZ_{\bs{Y}}&=\prod_{I,J=1}^N \prod_{s \in Y_I} {1\over \alpha_I -\alpha_J -\epsilon_1 v_{Y_J}(s) +\epsilon_2 \left( h_{Y_I}(s)+1 \right)} \\
&\qquad \qquad \times  \prod_{s\in Y_J}
{1\over \alpha_I -\alpha_J +\epsilon_1 \left(v_{Y_I}(s)+1\right){-} \epsilon_2 h_{Y_J}(s)},
\ea
\ee
with
\be  \sum_{i=1}^N \alpha_i=0, \ee
and 
\be a_i=\alpha_{i}-\alpha_{i+1} , \quad i=1,\cdots, N-1.\ee
The  four dimensional   $SU(N)$ Nekrasov-Shatashvili (NS) free energy is then defined by \cite{ns}
\begin{equation}
\label{fns4d}
F_{\rm NS}^{\rm inst}(\bs{a}, \hbar) =\ri \hbar \, \lim_{\epsilon_2
  \rightarrow 0} \, \epsilon_2 \log Z(\bs{a}; \ri \hbar , \epsilon_2).
\end{equation}
An important property of this free energy is that it is given as a
power series in $\Lambda$ which is expected to have a non-vanishing
radius of convergence in a certain range of values of $a$ and $\hbar$
around the semiclassical region $|\Lambda/a| \ll 1$\footnote{Recall
  that for $\hbar = 0$, in the electric frame both $a(u)$ and $a_D(u)$
  are convergent series of $\Lambda^2/u$ up to the monopole and dyon
  points.  The prepotential, which is related to $a,a_D$ by the
  special geometry relation, is thus also convergent series of
  $\Lambda/a$ up to these points.  The NS free energy is its smooth
  deformation which tends to enlarge the domain of convergence.}.
 
In order to make contact with the (modified) Mathieu equation, the relevant
gauge theory is $\CN=2$, $SU(2)$ SYM theory, hence we
have to consider \eqref{fns4d} with $N=2$. In this case the first few
terms read
\begin{equation}
  \label{inst} \ba F_{\rm NS}^{\rm inst}(a, \hbar)=-\frac{2 \Lambda
    ^4}{a^2+\hbar ^2}+ \frac{\Lambda ^8 \left(7 \hbar ^2-5
      a^2\right)}{\left(a^2+\hbar ^2\right)^3 \left(a^2+4 \hbar
      ^2\right)}+\CO\left(\Lambda ^{16}\right).\ea
\end{equation}
Once the NS free energy is known, the quantum
$A$ period $a(u,\hbar)$ can be obtained by inverting the quantum
Matone relation \cite{acdkv,lmn,francisco,bk,bkk,fm-wilson,sciarappa1} 
\begin{equation}
  \label{mmap}u= {a^2\over 8}-{\Lambda \over 8}
  \partial_{\Lambda}F_{\rm NS}^{\rm inst}(a, \hbar, \Lambda).
\end{equation}
This leads to a series expansion for $a(u, \hbar)$ in powers of
$\Lambda^{4}$ which is expected to converge in an appropriate range of the parameters $\Lambda$, $E$, $\hbar$ around the semiclassical region $|\Lambda^2/E| \ll 1$.  The quantum
$B$ period $a_D(u,\hbar)$ is then obtained from the quantum special geometry relation
\begin{equation}
  \ba a_D(a, \hbar)=\partial_{a}F_{\rm NS}(a, \hbar, \Lambda)=& 2
  \gamma (a,\hbar,\Lambda )+\partial_{a}F_{\rm NS}^{\rm inst}(a,
  \hbar, \Lambda)\\ 
  \ea
\end{equation}
where \cite{hm}
\begin{equation}
  \label{gammadef}
  \gamma (a,\hbar )=\frac{a}{2}  \log \left({\hbar ^2\over
        \Lambda^2}\right)-\frac{\pi \hbar }{4}-\frac{\ri
  \hbar }{2}  \left(\log \Gamma\left(1+\frac{\ri a}{\hbar }\right)-\log
    \Gamma\left(1-\frac{\ri a}{\hbar }\right)\right) \ ,
\end{equation}
and we replace $a$ by $a(u,\hbar)$.

We also note that it is possible to express the NS free energy via a
TBA system \cite{ns,kt2} (different from the one discussed in
section~\ref{sc:SW-TBA}). This TBA system however has a range of
validity/convergence which is smaller than the one of the instanton
calculus. For instance, the TBA breaks down if ${\rm Re}(a)\neq 0$,
while the instanton counting expression for $F_{\rm NS}^{\rm inst}$
\eqref{fns4d} is still well defined.

It was found in \cite{mirmor} that by expanding \eqref{inst},
\eqref{mmap} at small values of $\hbar$, it is possible to recover the
WKB periods \eqref{QP}. More precisely, at order $\hbar^{2n}$, one
finds an expansion in $\Lambda$ which agrees with the expansion of
$\Pi^{(n)}(E)$ at large $E$: 
\begin{equation}
  \label{limit}
  \ba
  &2 \pi \, a(u,\hbar)\longrightarrow \sum_{n=0}^\infty \hbar^{2n}
    \Pi_A^{(n)} (E=2u), \\ 
  & 2 \ri \, a_D(a(u,\hbar), \hbar)= 2 \ri \partial_a( F_{\rm NS}(a,
  \hbar))\Big|_{a= a(u,\hbar)}\longrightarrow 
\sum_{n=0}^\infty \hbar^{2n} \Pi_B^{(n)} (E=2u).
  \ea
\end{equation}
Therefore, the Bethe/gauge correspondence provides an analytic way to
resum the WKB periods period into well defined functions which are
exact in $\hbar$. We will denote these functions by
$\Pi_{A, B}^{\rm ex}(E, \hbar)$, and we will refer to them as ``exact"
quantum periods. In terms of the quantities that we have introduced,
they are given by
\be\label{abpe}  
\ba
\Pi^{\rm ex}_A(E, \hbar)&=2 \pi a(E,\hbar), \\
\Pi_B^{\rm ex}(E, \hbar) &={ 2 \ri }\partial_{a}F_{\rm NS}(a (E, \hbar),\hbar),
\ea
\ee
where we use the notation $a(E,\hbar):=a(u=E/2,\hbar)$.

It turns out that one can find the series expansion for $a(u, \hbar)$
by using elementary methods.  To do this, we use the WKB method, but we solve the Riccati
equation \eqref{riccati} perturbatively in $\Lambda$, i.e.\ we solve
\be
 Y^2(x)-\ri \hbar\frac{\rd Y(x)}{\rd x}=E- 2\Lambda^2 \cosh(x) 
\end{equation}
with an ansatz
\be
Y(x) = \sum_{n \ge 0} Y_n (x, \hbar, E) \Lambda^{2n}. 
\ee
Clearly, we should set 
\be
Y_0(x, \hbar, E)= {\sqrt{E}}, 
\ee
The equation for $Y_1(x, \hbar, E)$ is 
\be
2 Y_0 Y_1 -{\ri \hbar} {\rd Y_1 \over \rd x}= - 2 \cosh(x). 
\ee
The general solution to this equation is of the form 
\be
Y_1(x, \hbar, E)= -{4 {\sqrt{E}} \cosh(x) + 2 \ri \hbar \sinh(x) \over 4 E+ \hbar^2} + c \re^{-{2 \ri {\sqrt{E}} x\over \hbar}}. 
\ee
We note that the term involving the unknown coefficient leads to a
non-perturbative effect in $\hbar$. We will set it to zero to recover
the perturbative series.  The general term $Y_n$ satisfies
\be
2 {\sqrt{E}} Y_n(x, \hbar, E) -\ri \hbar {\rd Y_n \over \rd x}+\sum_{k=1}^{n-1} Y_k(x, \hbar, E) Y_{n-k}(x, \hbar, E)=0. 
\ee
This can be integrated order by order, setting to zero non-perturbative terms. We find in this way, 
\be
Y_2(x, \hbar, E)= -\frac{4 E^2+5 E \hbar ^2+\ri {\sqrt{E}}  \hbar  \left(8 E-\hbar ^2\right)
   \sinh (2 x)+E \left(4 E-5 \hbar ^2\right) \cosh (2 x)+\hbar ^4}{{\sqrt{E}}
    \left(E+\hbar ^2\right) \left(4 E+\hbar ^2\right)^2}. 
    \ee
    The functions $Y_n(x, \hbar, E)$ are complicated, but their integrals are slightly
    simpler. As it follows from (\ref{eq:PiA0}), we have to calculate
    \be
  \CI_n (E, \hbar)=  {1\over 2 \pi} \int_{-\pi}^\pi Y_n (\ri x, \hbar, E) \rd x. 
    \ee
As expected, only even terms contribute. We find, for example, 
 \be
   \ba
   \CI_2(E, \hbar)&=  -\frac{1}{4 E^{3/2}+{\sqrt{E}}  \hbar ^2}, \\
   \CI_4(E, \hbar)&=-\frac{60 E^2+35 E \hbar ^2+2 \hbar ^4}{4 E^{3/2} \left(E+\hbar
   ^2\right) \left(4 E+\hbar ^2\right)^3}, \\
   \CI_6(E, \hbar)&=-\frac{6720 E^5+18480E^4 \hbar ^2+15260 E^3 \hbar ^4+4705 E^2
   \hbar ^6+413 E \hbar ^8+18 \hbar ^{10}}{4 E^{5/2} \left(E+\hbar
   ^2\right)^2 \left(4 E+\hbar ^2\right)^5 \left(4 E+9 \hbar
   ^2\right)},
   \ea
   \ee
   and so on. Then, one finds
   \be
   \label{aseries}
  a(E, \hbar)= 2\left(\sqrt{E} + \sum_{m=1}^{\infty} \CI_{2m}(E, \hbar) \Lambda^{4m}\right).
   \ee
When $\hbar=0$, we recover the standard SW period \eqref{eq:PiA0}: 
   \be
2 \pi a (E, 0)=4 \pi{\sqrt{2+E}} ~_2 F_1\left( -{1\over 2}, {1\over 2}, 1; {4 \over 2+E}\right)={8} {\sqrt{2\Lambda^2+E}~} {\bf E}\left( {4\Lambda^2 \over {2\Lambda^2+E}} \right). 
   \ee
  At finite $\hbar, \Lambda$ we also find perfect agreement between \eqref{aseries} and the standard result of instanton calculus.

  We note that the integrals above can be calculated as residues, since 
  \be
  \CI_n (E, \hbar)=\oint_{|X|=1} Y_n(X, \hbar, E) {\rd X \over  X} ={\rm Res}_{X=0} { Y_n(X, \hbar, E) \over X}, 
  \ee
  where
  \be
  X=\re^x. 
  \ee
  In fact, it is more convenient to solve the differential equation
  directly in the $X$ variable, since everything is algebraic, i.e.\ it
  is better to solve
  \be 2 \sqrt{E} Y_n(X, \hbar, E) -\ri \hbar X {\rd Y_n \over \rd X
  }(X, \hbar, E)+\sum_{k=1}^{n-1} Y_k(X, \hbar, E) Y_{n-k}(X, \hbar,
  E)=0.  \ee

It turns out that the function $Y(\ri x, \hbar, E)$ can be calculated exactly in terms of Mathieu functions. To see this, we note that $\CY(x, \hbar, E)=Y(\ri x, \hbar, E)$ satisfies the 
Riccati equation 
\be
\CY^2(x, \hbar, E) -\hbar {\rd \CY \over \rd x}(x, \hbar, E)= E- 2 \Lambda^2 \cos(x), 
\ee
which is the Mathieu equation with imaginary Planck constant. The solution to this equation is 
\be
\CY(x, \hbar, E)= -\hbar {\rd \over \rd x} \log \left\{ S \left(-{4 E \over \hbar^2}, -{4 \Lambda^2 \over\hbar^2}, {x\over 2}\right)+c C \left(-{4 E \over \hbar^2}, -{4 \Lambda^2 \over\hbar^2}, {x\over 2}\right)\right\}, 
\ee
where $c$ is an integration constant and $S(\alpha, q, x)$, $C(\alpha,q,x)$ are the odd (even) Mathieu functions, respectively. Since we want 
\be
\CY(x, \hbar, E) \approx {\sqrt{E}}
\ee
as $\Lambda \rightarrow 0$, we find that this leads to 
\be
c=\pm \ri
\ee
 (where the sign depends on the branch cut of the square root). We eventually find 
\be
\label{inta}
\Pi^{\rm ex}_A (E, \hbar) =  {2 }\int_{-\pi}^{\pi} \CY(x) \rd x = -{2 \hbar } \log {S \left(-{4 E \over \hbar^2}, -{4 \Lambda^2 \over\hbar^2}, {\pi\over 2}\right)+c C \left(-{4 E \over \hbar^2}, -{4 \Lambda^2 \over\hbar^2}, {\pi\over 2}\right) \over -S \left(-{4 E \over \hbar^2}, -{4 \Lambda^2 \over\hbar^2}, {\pi\over 2}\right)+c C \left(-{4 E \over \hbar^2}, -{4 \Lambda^2 \over\hbar^2}, {\pi\over 2}\right)}. 
\ee
Note that 
\be
\label{a-char}
\Pi^{\rm ex}_A(E, \hbar) =2 \pi  {\ri \hbar } \,\nu\left( -{4 E \over \hbar^2}, -{4 \Lambda^2 \over\hbar^2} \right), 
\ee
where $\nu$ is the characteristic exponent of the Mathieu equation (this relation has been noted in the context of the Mathieu equation in e.g. \cite{he-miao, kpt}). 
The advantage of the expressions \eqref{inta}, (\ref{a-char}) is that they make sense for values of $E$ for which the $\Lambda$ expansion \eqref{aseries} does not converge, so they extend 
(\ref{aseries}) to a larger domain. 

Let us consider some numerical examples. When $E=100$, $\hbar=3 \pi$ and $\Lambda=1$, we can evaluate the series (\ref{aseries}) by truncating it up to order $\Lambda^{28}$, and we find:
\be
{1\over 4 \pi } \Pi^{\rm ex}_A(100, 3 \pi)=-9.9997954179096157891757...
\ee
This is precisely what is also obtained from (\ref{inta}) and (\ref{a-char}). At the same time, by using (\ref{inta}) we can go all the way to $E=0$, where (\ref{aseries}) cannot be used. We find, for example, 
\be
 {1\over  4 \pi }\Pi^{\rm ex}_A(0, 3 \pi)=0.1501122164563802133431995... \ri. 
\ee
This procedure for evaluating the value of $\Pi^{\rm ex}_A$ at $E=0$ seems to be well-defined for sufficiently large $\hbar \Lambda^{-2}$ (e.g. $\hbar \Lambda^{-2} \ge 1$ works). 

We conclude that the ``exact" quantum $A$ period can be computed either
by the expression given by instanton calculus (or equivalently, by the
closely related series (\ref{aseries})), or by the expression
(\ref{a-char}) involving the Mathieu characteristic exponent
(\ref{a-char}). When these two expressions are both well-defined, they
agree, but (\ref{a-char}) has a larger range of validity. In the case
of the quantum $B$ period, it might be possible to obtain an alternative
expression to the one in (\ref{abpe}), in terms of infinite Hill
determinants, by using results in \cite{gutz1}.

\subsection{Comparison to Borel resummation}\label{bo}

We now have two different approaches to the calculation of (resummed)
quantum periods: on the one hand, we have the Borel resummation of the
all-orders WKB expansion in $\hbar^2$, which is also calculated by the
TBA equations of section \ref{tbagmn}. On the other hand, instanton
calculus gives a different resummation, based on a convergent
expansion in $\Lambda$, as a function of $\hbar$. An obvious question
is: what is the precise relation between these two resummations? Since
both lead to the same asymptotic expansion in powers of $\hbar$, we
expect that they will differ in non-perturbative effects. In this
section we will address this issue.  Results along these lines have
been previously obtained in \cite{kpt,ashok}. For simplicity, we will
restrict ourselves to the case in which $\hbar>0$ and $u$ is real.

Let us first consider the weak coupling region $\CW$. Here, the
all-orders WKB quantum $A$ period is Borel summable for $\hbar$ real,
and we find that its Borel sum agrees well with $2 \pi a(u, \hbar)$
obtained by inverting \eqref{mmap} or with the solution \eqref{inta}
to the Riccati equation, i.e.
\be
s(\Pi_A)(E,\hbar)= \Pi_A^{\rm ex}( E,\hbar), \qquad |u|>1, \quad \hbar>0. 
\ee
We illustrate this in Table~\ref{tb:BorelSum-A} where we compare the
Borel sum of the WKB quantum $A$ period at $u=4$ and $\hbar=1$, with
increasing number of corrections, to the result of instanton
calculus. They agree with almost all the stabilised digits (27 of
them).

\begin{table}
  \centering
  \begin{tabular}{cc}\toprule
    terms included & Borel sum \\\midrule
    190 & 35.40661948105291481767982565157 \\
    191 & \underline{35.40661948105291481767982565}207\\\midrule
    gauge theory & 35.40661948105291481767982564492 \\\bottomrule
  \end{tabular}
  \caption{Borel sum of quantum $A$ period at $u=4$ and
    $\hbar=1$. Underlined are stabilised digits.}
  \label{tb:BorelSum-A}
\end{table}

The quantum $B$ period, on the other hand, is not Borel summable along
the real axis in the weak coupling region. Nevertheless, we can make
the following observation. The ``exact" quantum $B$ period is given by
\begin{equation}
  \label{adgauge} 2 \ri a_D(u, \hbar)=4 \ri \gamma
  (a,\hbar )+2 \ri \partial_{a}F_{\rm NS}^{\rm inst}(a, \hbar)
\end{equation}
where $\gamma$ is defined in \eqref{gammadef} and has the following
asymptotic expansion for large $a/\hbar$:
\begin{equation}
  \label{asympt}
  \gamma(a, \hbar)\sim a \left( \log \left( {a \over \Lambda}
    \right)-1\right)+a \sum_{n \ge 1} {(-1)^n B_{2n} \over 2n (2n-1)}
  \left({\hbar \over a}\right)^{2n}.  
\end{equation}
The series in $\hbar/a$ in the r.h.s. is not Borel summable along the
positive real axis. More precisely, let us consider the following
formal power series:
\begin{equation}
  \varphi(z)= \sum_{n \ge0} c_n z^{2n}, \qquad c_{n-1}={ (-1)^n B_{2n}
    \over 2n (2n-1)}.   
\end{equation}
A little numerical experimentation shows that the lateral
resummations of this series along the positive real axis are given by
\begin{equation}
  \label{disc-gamma}
  s_{\pm}(\varphi)(z)= f(z) \pm {\ri \over 2z} \log \left(1- \re^{-2\pi/z} \right), 
\end{equation}
where 
\begin{equation}
  f(z)= {1\over z^2} \left( \log(z)+1 \right)-{\pi \over 4z}  -{\ri
    \over 2 z} \log {\Gamma( 1+\ri/z) \over \Gamma(1-\ri /z) }. 
\end{equation}
(A similar series has been considered in \cite{hm}). 
\begin{figure}
  \centering
  \includegraphics[width=0.45\linewidth]{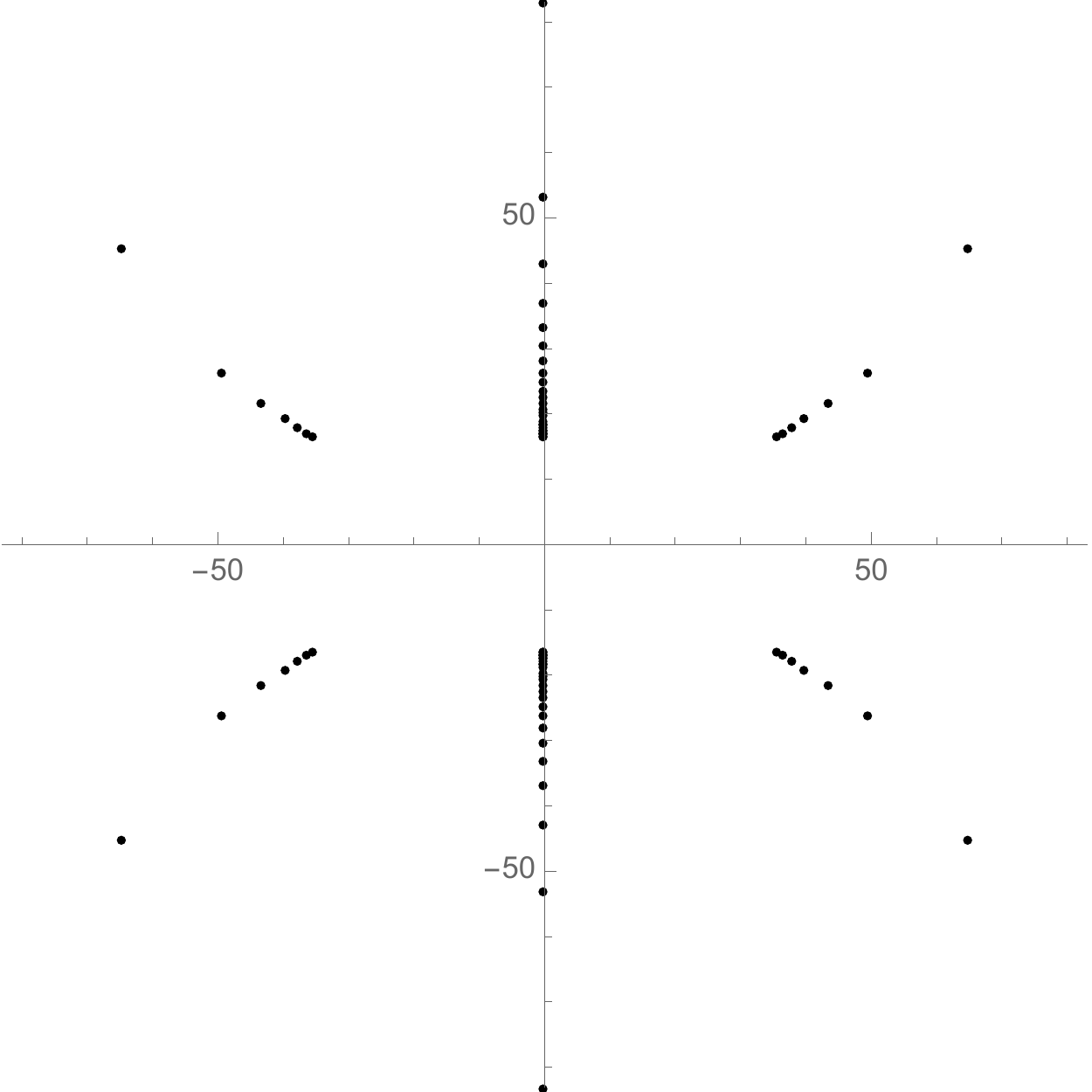}
  \caption{Poles of the Borel--Pad\'e transform of the ``reduced" series
    $\Pi_B^{\rm red}$ at $u=4$.}
  \label{fg:Borel-aDformal}
\end{figure}

The above analysis suggests that the non-Borel summability of the
sequence $\Pi_B^{(n)}$ along the positive real axis in the weak
coupling region is due to the asymptotic series appearing in
$\gamma(a,\hbar)$. In view of (\ref{disc-gamma}), this leads to the
right discontinuity across the positive real axis:
\begin{equation}
  -4 \hbar \log (1 -\re^{-4 \pi a/\hbar}). 
\end{equation}
This suggests that the ``reduced" formal power series
\begin{equation}\label{piBformal}
  \Pi^{\text{red}}_B(E,\hbar):=
  \Pi_B(E,\hbar)-4 \ri \Pi_A (E,\hbar)
  \sum_{n \ge 1} {(-1)^n B_{2n} \over 2n (2n-1)}
  \left( {\hbar \over 2\Pi_A(E,\hbar)}\right)^{2n},
\end{equation}
where we subtract the non-Borel summable series in the function
$\gamma(a, \hbar)$, is actually Borel summable along the positive real
axis. We verified numerically that this indeed is true, as can be seen
from the Borel plane plot at $u=4$ given in
Figure~\ref{fg:Borel-aDformal}. In fact the Borel sum of
$\Pi^{\text{red}}_B(E,\hbar)$
agrees with the gauge theory calculation in which the contribution of
$\gamma(a,\hbar)$ has been removed; in other words,
\begin{equation}\label{ared}
  s\left( \Pi^{\text{red}}_B \right)(E,\hbar)=2\ri a_D^{\text{red}}(u,\hbar):= 4\ri
  a\(\log\(\frac{a}{\Lambda}\)-1\)
  +2\ri\frac{\pd F^{\text{inst}}_{\text{NS}}(a,\hbar)}{\pd a} \ .
\end{equation}
We illustrate this in Table~\ref{tb:BorelSum-B}, where both the Borel
sum of $\Pi^{\text{red}}_B(E,\hbar)$ and
$2\ri a_D^{\text{red}}(u,\hbar)$ are evaluated at $u=4$ and
$\hbar=1$. We find that all stabilised digits are in agreement (26 of
them)\footnote{We also notice that the exact $B$ period
  $\Pi_B^{\text{ex}}(E,\hbar)$ with $E>2$ agrees with the average of
  lateral Borel resummations of the quantum $B$ period
  $\Pi_B(E,\hbar)$.}.

We can now see a clear difference between the TBA equations of
\cite{ns} and the TBA equations of \cite{gmn}. The TBA equations of
\cite{ns} compute the Borel-summable part of the $B$ period, where we
have removed the perturbative contribution due to the $\gamma$
function, i.e.~they compute \eqref{ared}, which is the Borel
resummation of $\Pi_B^{\rm red}$ in \eqref{piBformal}.  On the
contrary, the conformal limit of the GMN TBA equations computes the
Borel resummation of the {\it full} quantum $B$ period $\Pi_B(E,\hbar)$,
including the perturbative $\gamma$ function. Since the latter is not
Borel summable, the corresponding TBA has discontinuities, as discussed in
section~\ref{sc:SW-TBA}.

\begin{table}
  \centering
  \begin{tabular}{ccl}\toprule
    & terms & \\\midrule
    \multirow{2}{*}{$\Pi^{\text{red}}_B$}
    &159 & 16.474810551500808917635392219 \\
    &161 & \underline{16.474810551500808917635392}368 \\\midrule
    \multirow{2}{*}{$2\ri a_D^{\text{red}}$}
    &14 & 16.47481055150080891763539232909\\
    &15 & \underline{16.474810551500808917635392329}20 \\\bottomrule
  \end{tabular}
  \caption{Comparison between the Borel sum of the ``reduced" quantum $B$ period
    $\Pi^{\text{red}}_B$ at $u=4$ and $\hbar=1$ (with
    increasing number of terms), and the exact, ``reduced" period
    $2\ri a_D^{\text{red}}$ defined in (\ref{ared}) (with
    increasing number of instanton corrections). Underlined are
    stabilised digits.}
  \label{tb:BorelSum-B}
\end{table}

\begin{table}
  \centering
  \begin{tabular}{cl}\toprule
    terms & first line of r.h.s. of \eqref{eq:Pi-ex}\\\midrule
    191 & $806502.11499751621351505261143$\\
    193 & $\underline{806502.11499751621351505261}260$ \\\midrule
    l.h.s. of \eqref{eq:Pi-ex}
          & $806502.11499751621351505261016$\\
    \bottomrule                  
  \end{tabular}
  \caption{Numerical verification of the characteristic exponent
    formula \eqref{eq:Pi-ex} at $u=0$ and $\hbar=1/4$, 
    with increasing number of terms for the lateral Borel
    resummations on the r.h.s.. All the stabilised digits (underlined)
    are in agreement.}
  \label{tb:Pi-ex}
\end{table}

In the strong coupling region we have the
following relation between the ``exact" quantum $A$ period and lateral 
Borel resummations of quantum periods \cite{kpt}
\begin{equation}\label{eq:Pi-ex}
\begin{aligned}
  2 \cosh(\Pi_A^{\text{ex}}/(2\hbar))
  &= \re^{\frac{1}{2\hbar}s_+(\Pi_A)} + \re^{\frac{1}{2\hbar}s_+(\Pi_A+2\Pi_B)}
  + \re^{-\frac{1}{2\hbar}s_+(\Pi_A+2\Pi_B)} \\
  &=\re^{\frac{1}{2\hbar}s_-(\Pi_A)} + \re^{-\frac{1}{2\hbar}s_-(\Pi_A)} +
    \re^{\frac{1}{2\hbar}s_-(\Pi_A+2\Pi_B)} \ .
  \end{aligned}
\end{equation}
Numerical evidence for this relation is presented in
Table~\ref{tb:Pi-ex} for the first line of the formula, evaluated at
$u=0$ and $\hbar=1/4$. 
The second line of \eqref{eq:Pi-ex} can be
derived from the Stokes automorphism of quantum periods discussed in
section \ref{sc:SW-TBA}.
In the strong coupling region, $\Pi_D=\Pi_B+\Pi_A$ is Borel summable
along the positive real axis, while both the Borel resummations of
$\Pi_B$ and $\Pi_A$ have discontinuities across the positive real
axis. The  discontinuities for the $B$, and $A$ periods
are:
\begin{equation}
  \begin{aligned}
    &s_+(\Pi_B) - s_-(\Pi_B) = 2\hbar \log
    (1+\re^{-\frac{1}{\hbar}s(\Pi_D)})\ ,\\
    &s_+(\Pi_A) - s_-(\Pi_A) = -2\hbar \log
    (1+\re^{-\frac{1}{\hbar}s(\Pi_D)}) \ .
  \end{aligned}
\end{equation}
Starting from the first line on the right hand side of
\eqref{eq:Pi-ex}, and applying the discontinuity formulae, we
immediately get the second line
\begin{align}
  &\text{first line}\nn
  &=
    \re^{\frac{1}{2\hbar}s_+(\Pi_A)} + \re^{\frac{1}{2\hbar}s_+(\Pi_A+2\Pi_B)}
    + \re^{-\frac{1}{2\hbar}s_+(\Pi_A+2\Pi_B)}
    \nn
  &=
    \re^{\frac{1}{2\hbar}s_-(\Pi_A)}\(1+\re^{-\frac{1}{\hbar}s(\Pi_D)}\)^{-1}
    + \re^{\frac{1}{2\hbar}s_-(\Pi_A+2\Pi_B)}\(1+\re^{-\frac{1}{\hbar}s(\Pi_D)}\)
    +
    \re^{-\frac{1}{2\hbar}s_-(\Pi_A+2\Pi_B)}
    \(1+\re^{-\frac{1}{\hbar}s(\Pi_D)}\)^{-1} 
    \nn
  &=
    \(\re^{\frac{1}{2\hbar}s_-(\Pi_A)}+\re^{-\frac{1}{2\hbar}s_-(\Pi_A+2\Pi_B)}\)
    \(1+\re^{-\frac{1}{\hbar}s(\Pi_D)}\)^{-1} +
    \re^{\frac{1}{2\hbar}s_-(\Pi_A+2\Pi_B)}+\re^{-\frac{1}{2\hbar}s_-(\Pi_A)}
    \nn
  &=
    \re^{\frac{1}{2\hbar}s_-(\Pi_A)}+\re^{-\frac{1}{2\hbar}s_-(\Pi_A)}
    +\re^{\frac{1}{2\hbar}s_-(\Pi_A+2\Pi_B)}  = \text{second line} \ .
\end{align}
On the other hand, to derive a similar result for the ``exact''
quantum $B$ period we can use results on the Fredholm determinant of
the modified Mathieu equation, which we present in
section~\ref{tssst}. Eq.\eqref{bspd2} then together \eqref{eq:Pi-ex}
imply that
\begin{equation}\label{eq:Pi-ex-2}
\begin{aligned}
  2\sinh(\Pi^{\text{ex}}_B/(2\hbar))
  &= \re^{\frac{1}{2\hbar} s_+(\Pi_B)}-\re^{-\frac{1}{2\hbar}
    s_+(\Pi_B)}
    - \re^{-\frac{1}{2\hbar} s_+(2\Pi_A+3\Pi_B)}, \\
  &= \re^{\frac{1}{2\hbar} s_-(\Pi_B)}-\re^{-\frac{1}{2\hbar} s_-(\Pi_B)}
    +\re^{-\frac{1}{2\hbar} s_-(2\Pi_A+\Pi_B)} \ ,
  \end{aligned}
\end{equation}
and we have tested these identities numerically to very high
precision.

Summarizing, in the weak coupling region the all-orders WKB quantum $A$
period and the $B$ period (once the gamma function is subtracted) are
Borel summable. Their Borel sums agree with the gauge theory
expressions of section~\ref{ic}. In the strong coupling region, the $A$ and
$B$ periods are not Borel summable, although their lateral Borel resummations can
be related to the exact quantum $A$ period via \eqref{eq:Pi-ex}. We
finally note that the combination $\Pi_D=\Pi_A+\Pi_B$ is Borel
summable in the strong coupling region (only). However, its Borel sum
does not agree with the gauge theory expression of section~\eqref{ic},
namely
\begin{equation}
s(\Pi_D)\neq  \Pi_A^{\rm ex}+\Pi_B^{\rm ex}= 2 \pi a+2 \ri a_D, 
\end{equation}
and one should include additional non-perturbative corrections.  We
will find the correct formula at the end of section~\ref{comp-zamo}.

\section{The Fredholm determinant from topological string theory} \label{tssst}

Let $\mO$ be an operator on $L^2(\IR)$ such that $\mO^{-1}$ is of trace class. Then, $\mO$ has a discrete spectrum $\{E_n\}_{n\geq 0}$, and its Fredholm determinant 
\be
\label{fdm} \Xi(E)={\rm det}\left(1-E \mO^{-1} \right) =\prod_{n\geq0}\left({1-{E \over E_n}}\right)
 \ee
is an entire function of $E$ whose zeros give the spectrum of $\mO$: $\Xi(E_n)=0$ (see e.g. \cite{simon-paper} for these and other properties of Fredholm determinants). 

The Fredholm determinant contains very rich information about the spectral properties of $\mO$. For example, the spectral traces, defined as
 \be\label{trdef} Z_\ell=\sum_{n \geq 0}{1\over E_n^\ell}, 
 \ee
can be computed by expanding the spectral determinant around $E=0$. Indeed, we have
  \be 
  \Xi(E)=\sum_{N\geq 0} (-E)^N Z(N,\hbar) 
  \ee
where
\be
\label{conjclasses}
Z(N, \hbar) =\sum_{\{ m_\ell \}} {}^{'}\prod_\ell   {(-1)^{(\ell-1)m_\ell } Z_\ell^{m_\ell} \over m_\ell! \ell^{m_\ell}},
\ee
and the $ {}^{'}$ means that the sum is over the integers $m_\ell$ satisfying the constraint
\be
\label{Ncons}
\sum_\ell \ell m_\ell=N.
\ee
From the quantities $Z(N, \hbar)$ (which were called {\it fermionic
  spectral traces} in \cite{ghm}) one can extract the conventional
spectral traces (\ref{trdef}).

Although Fredholm determinants are central objects in spectral theory,
it is difficult to obtain explicit expressions for them. It is easy to
show (see for instance \cite{lst}) that the inverse of the modified
Mathieu operator \eqref{eq:schrodinger} is of trace class. Therefore,
the Fredholm determinant is well-defined, and it is an interesting
question to find an explicit, closed form expression for this
quantity.

In recent years it was discovered \cite{km, hw, cgm8, ghm, cgm} that, by
using topological string tools, it is possible to obtain explicitly
expression for Fredholm determinants of operators arising in the
context to quantum mirror curves.  We will refer to this relationship
as the TS/ST correspondence. As explained in section~\ref{deri}, the
modified Mathieu operator \eqref{eq:schrodinger} can be related, upon
a suitable limiting procedure, to the quantum mirror curve of local
$\IF_0$. Therefore we can study \eqref{fdm} within the context of
topological string theory and in particular, by using \cite{ghm, gm3},
we can deduce an explicit, closed form expression for the Fredholm
determinant of the modified Mathieu operator.  We will first state the
main result and then explain how to derive it within topological
string theory. We also present several independent tests of our
result, including an interesting connection to the TBA system of
\cite{post-zamo}.

\subsection{A closed formula and its derivation}
\label{deri}

By using the approach of \cite{ghm,gm3} we find the following expression 
for the spectral determinant of \eqref{eq:schrodinger}\footnote{While presenting these 
results at the conference {\it Irregular singularities in Quantum Field Theory} (http://irregular.rd.ciencias.ulisboa.pt/conference),   	
S. Lukyanov informed us that he had independently derived this result \cite{PCL} by using completely different methods. }
\be 
\label{sd}\Xi(E)=A (\hbar) \left( \sinh\left(\frac{ \Pi_A^{\rm ex}(E, \hbar)}{2 \hbar }\right) \right)^{-1} \cosh\left({1 \over 2\hbar} \Pi_B^{\rm ex}(E, \hbar)\right), 
\ee
where $\Pi_{A,B}^{\rm ex} (E, \hbar)$ are given by (\ref{abpe}) and $A(\hbar) $ is an $u$-independent constant which can be fixed from $\Xi(0)=1$, namely, 
 \be\label{ainst} A(\hbar) =  \sinh\left(\frac{ \Pi_A^{\rm ex}(0, \hbar)}{2 \hbar }\right)  \left(\cosh\left({1 \over 2\hbar} \Pi_B^{\rm ex}(0, \hbar)\right)\right)^{-1}.\ee
From this expression we can read off explicit formulae for the spectral traces. We find for instance 
\be \label{tr1} Z_1=-{1\over 2 \hbar}\left(\partial_E\Pi_B^{\rm ex}(0, \hbar) \tanh \left( {\Pi_B^{\rm ex}(0, \hbar)}\over 2 \hbar\right)-  \coth \left(\frac{\Pi_A^{\rm ex}(0, \hbar)}{2 \hbar }\right)   \partial_E \Pi_A^{\rm ex}(0, \hbar)\right)      ,\ee
as well as
\be \ba \label{tr2} 4 \hbar^2 Z_2= & 2 \hbar \partial_E^2\Pi_A^{\rm ex}(0,\hbar) \coth \left(\frac{\Pi_A^{\rm ex}(0,\hbar)}{2 \hbar }\right)-(\partial_E\Pi_A^{\rm ex}(0,\hbar))^2 \text{csch}^2\left(\frac{\Pi_A^{\rm ex}(0,\hbar)}{2 \hbar }\right)\\
&-\text{sech}^2\left(\frac{\Pi_B^{\rm ex}(0,\hbar)}{2 \hbar }\right) \left(\hbar  \partial_E^2 \Pi_B^{\rm ex}(0,\hbar) \sinh \left(\frac{\Pi_B^{\rm ex}(0,\hbar)}{\hbar }\right)
+(\partial_E \Pi_B^{\rm ex}(0,\hbar))^2\right). 
\ea
 \ee
 Note that in order to calculate $\Pi_A^{\rm ex}(0,\hbar)$ we have to use (\ref{inta}) or (\ref{a-char}), so the above formula tests as well the analytic continuation of instanton calculus beyond the semiclassical region\footnote{In evaluating the derivative of the periods w.r.t.~$E$, we used the quantum Matone relation (\ref{mmap}) and instanton 
 calculus. }.  

The explicit formula (\ref{sd}) can be extended to the family of operators considered in \cite{gm3}, see Appendix \ref{asun} for more details.

Let us now explain how to
derive \eqref{sd} from the TS/ST correspondence of \cite{ghm,cgm}. The relevant CY geometry is the
canonical bundle over $\IF_0$, also known as resolved $Y^{2,0}$
singularity.  The corresponding quantum mirror curve is
\begin{equation}
\mO-\kappa=(R \Lambda)^2(\re^{\mathsf x}+\re^{-\mathsf
    x})-\kappa+\re^{\mathsf p}+\re^{-\mathsf p},  \quad [\mathsf x,
  \mathsf p]=\ri  \hbar.
\end{equation}
According to the TS/ST correspondence we have
\begin{equation}
  \label{tsst} \det (1-\kappa \mO^{-1})=\sum_{ { m}
    \in \IZ} \exp\left[\mJ ( {\mu+\ri \pi+2 \pi \ri m},R\Lambda,
    \hbar)\right], \quad \kappa=\re^{\mu},
\end{equation}
where $\mJ ( {\mu},\xi, \hbar)$ is the grand potential of the resolved $Y^{2,0}$ singularity as defined in \cite{gm3}, section~5.1.
The expression  \eqref{sd}  is obtained by implementing  the geometric engineering limit \cite{selfdual,kkv} in \eqref{tsst}. 
More precisely, we consider the limit 
\be\label{4dlimit} p\to R p, \quad\hbar\to R\hbar, \quad \kappa=2+E R^2+\mathcal{O}(R^3), \quad R\to 0.\ee
This has to be done carefully since both sides of \eqref{tsst} diverge, therefore they need to be properly regularized. 
For that it is convenient to study the trace of the resolvent
\be 
G_\mO (\kappa)={\rd \over \rd \kappa} \log \det (1-\kappa \mO^{-1})={\rm Tr} \left({1\over \kappa- \mO}\right), 
 \ee
rather than the spectral determinant. It is easy to see that in the limit \eqref{4dlimit} one has
\be  G_\mO (\kappa) \to {1\over R^2} G(E), \qquad G(E)= \Tr\left({1\over E-\mH}\right),
\ee
where $\mH$ is the modified Mathieu operator in (\ref{eq:schrodinger}). 
Likewise the limit \eqref{4dlimit} can be implemented on the r.h.s~of \eqref{tsst} in a quite 
straightforward way by following \cite{gm3}, section~5.2 and by using the identity (3.9) in \cite{bgt}. 
The overall divergent piece $R^{-2}$ cancels and we find
\be 
\ba 
G(E)=-{\partial_E a \over \hbar}\left(\partial_a^2 F_{\rm NS}\tanh\left({\partial_a F_{\rm NS}\over \hbar}\right)+\pi \coth\left({a\pi\over \hbar}\right)\right).
\ea\ee
 By integrating w.r.t.~$E$ we obtain \eqref{sd}, where $A(\hbar)$ is an integration constant.

\subsection{Tests of our formula}
We will now test the expression (\ref{sd}) in several ways. 

A first simple test is that the zeros of $\Xi(E)$ give the correct spectrum of the modified Mathieu operator. This should be expected 
from the general results of \cite{gm3}, but it is instructive to check it explicitly. 
The zeros correspond to the vanishing of the $\cosh$ in (\ref{sd}), and by using 
(\ref{abpe}) we find,
\be 
\label{eqref} \partial_{a}F_{\rm NS}(a,\hbar)= \hbar  \pi  \left( n+{1\over 2} \right), 
\ee
which is the exact quantization condition obtained from the conjecture in \cite{ns}, subsequently proved in \cite{kt2}.

 Another check can be obtained by comparing our expression to the asymptotics of Fredholm determinants obtained in \cite{voros,voros-zq,voros-quartic} 
 by using the all-orders WKB method. This asymptotic expansion is valid when $E<0$, where the Fredholm determinant is not oscillatory. In order to write down 
 the asymptotics, we need some ingredients. Let 
\be
G(E) =\sum_{n =0}^\infty {1\over E-E_n}
\ee
be the trace of the resolvent, and 
\be
T(E)=  {1\over 2} \int_\IR {\rd x \over {\sqrt{2\Lambda^2 \cosh(x)-E}}} 
\ee
the transit time. We also need the formal power series in $\hbar$ 
\be
\label{aewkb}
b(E) =  \exp \left\{  \ri \sum_{n\ge 1} \hbar^{2n-1} \int_\IR p_{2n} (x, E) \rd x \right\}, 
\ee
where the functions $p_{2n}(x,E)$ are the ones appearing in the solution to the all-orders WKB method in \eqref{ricwkb}. 
Let us now define $\kappa=-E$, which will be taken to be positive. It is convenient to introduce the functions
\be
\widetilde G(\kappa)= G(-\kappa), \qquad \widetilde \Xi(\kappa)= \Xi(-\kappa), \qquad \widetilde b(\kappa) =b(-\kappa), \qquad \widetilde T(\kappa)= T(-\kappa).
\ee
Then, one has the following small $\hbar$ asymptotics,  
 \be
 \label{xi-as}
\widetilde \Xi(\kappa)\sim  \widetilde b(\kappa) \exp\left\{ {1\over \hbar} \int_0^\kappa \widetilde T(\kappa') \rd \kappa' - \int_0^\infty  \left( \widetilde G(\kappa') +{1\over \hbar} \widetilde T(\kappa') \right) \rd \kappa' \right\}.   
 \ee
The second term in the exponent is independent of $\kappa$ but depends on $\hbar$. We note that all the integrals involved in this expression are well 
defined precisely because 
$E$ is negative. The very first terms in the asymptotics can be easily worked out, and one finds\footnote{This agrees with an unpublished calculation of 
Y. Hatsuda, who obtained the same result by considering the semiclassical expansion of the spectral traces.} 
\be
\label{wkb-sd}
\log\Xi(-\kappa) \sim - {\ri \over 2 \hbar} \Pi_B^{(0)}(\kappa)+{\ri \hbar \over 2}\Pi_B^{(1)}(\kappa)+ \cdots,  \qquad \kappa>0, 
\ee
up to $\kappa$-independent terms. Note that the sign in the subleading correction is the opposite one to what one finds in the WKB expansion of the quantum $B$ period. 

 Let us now compare this result to the exact expression for the spectral determinant \eqref{sd}, which can be written as
 \be
 \label{alt-sd}
 \Xi(E)=A (\hbar) \left( \sinh\left(\frac{\pi a}{\hbar }\right) \right)^{-1} \cos\left( {1\over \hbar} {\partial F_{\rm NS} \over \partial a} \right).\
 \ee
From the explicit expression (\ref{aseries}) it is easy to see that,
when $E$ is negative, $a$ is purely imaginary. More precisely, one has
\be
\label{aminus}
a(-\kappa, \hbar)= \ri \alpha, \qquad \alpha= a(\kappa, \ri\hbar),  
\ee
and we take $\alpha>0$ for definiteness. In addition,
\be
\label{fminus}
\partial_a F^{\rm inst} _{\rm NS} (a, \hbar)= \ri \partial_\alpha F^{\rm inst}_{\rm NS} (\alpha, \ri\hbar).  
\ee
By using the explicit expression (\ref{gammadef}) and standard
identities for the $\Gamma$ function, we find
\be
\ba
{\Xi(-\kappa)\over A(\hbar)}&={1\over 2 \pi} \Gamma\left( 1+{\alpha \over \hbar} \right)  \Gamma\left( {\alpha \over \hbar} \right) \exp \left[ {\alpha \over \hbar} \log\left( {\hbar^2 \over \Lambda^2} \right) + {1\over \hbar} \partial_\alpha F^{\rm inst}_{\rm NS} (\alpha, \ri\hbar) \right]
\\& +{1\over 2 \pi} \Gamma\left(1-{\alpha \over \hbar} \right)  \Gamma\left(-{\alpha \over \hbar} \right) \exp \left[ -{\alpha \over \hbar} \log\left( {\hbar^2 \over \Lambda^2} \right) - {1\over \hbar} \partial_\alpha F^{\rm inst}_{\rm NS} (\alpha, \ri\hbar) \right]. 
\ea
\ee
The term in the second line gives an exponentially small correction to
the leading asymptotics. The small $\hbar$ asymptotics of the quantity
in the first line is given by 
\be
\ba
&\log \left[ {1\over 2 \pi} \Gamma\left( 1+{\alpha \over \hbar} \right)  \Gamma\left( {\alpha \over \hbar} \right) \right]+ {\alpha \over \hbar} \log\left( {\hbar^2 \over \Lambda^2} \right) + {1\over \hbar} \partial_\alpha F^{\rm inst}_{\rm NS} (\alpha, \ri\hbar) 
\\&\sim  -{\ri \over 2 \hbar} \sum_{n \ge 0} (-1)^n \Pi_B ^{(n)}(\kappa) \hbar^{2n}.  
\ea
\ee
We have used that, due to (\ref{aminus}) and (\ref{fminus}), the
quantum period is evaluated at $-E$, where $E=-\kappa<0$, and we have
to change  $\hbar\rightarrow \ri\hbar$. The result is in agreement with
the WKB asymptotics obtained in (\ref{wkb-sd}).

A more precise test of (\ref{sd}) can be made by comparing the analytical formulae for the spectral traces with numerical results. These are obtained 
by calculating the spectrum of $\mH$ with standard techniques. An example of such a comparison is shown in Table \ref{tbx1}. 
\begin{table}[h!] 
\centering
   \begin{tabular}{l l l}
  \\
 $N_b$& $Z_2$  \\
\hline
2 &  \underline{0.004794786}11468342466\\
 4 & \underline{0.004794786073913}81196\\ 
 6 &                 0.00479478607391375025           \\  
 \hline
Num &  0.00479478607391375025    \\
\end{tabular}    \\
\caption{ The second spectral trace $Z_2$ as computed from  \eqref{tr2} for $\hbar=3\pi$ and $\Lambda=1$. The number $N_b$ means that we truncate the series \eqref{inst}  at order $\Lambda^{4 N_b}$. The last line gives the numerical result obtained from the spectrum of $\mH$. }
   \label{tbx1}
 \end{table}

We finally note that $\Xi(E)$ is an entire function of $E$. In particular, the would-be singularities due to the denominator of (\ref{alt-sd}) or to 
the Gamma functions in (\ref{gammadef}) must cancel in the end. This leads in turn to constraints on the form of the singularities of $F_{\rm NS}^{\rm inst}$, 
which might be testable against the results in \cite{gorsky-bands} (see also \cite{beccaria}). 

\subsection{Comparison to Zamolodchikov's TBA equation}

\label{comp-zamo}
An additional test of our formula (\ref{sd}) comes from a comparison
with \cite{post-zamo}.  Inspired by the ODE/IM correspondence
\cite{dt,ddt}, Zamolodchikov found in \cite{post-zamo} a TBA equation
which computes precisely the spectral determinant \eqref{sd}. Let us
state the main result of \cite{post-zamo}, referring to Appendix
\ref{appendixzamo} for more details. Let $\epsilon(\theta, P)$ be a
solution of the TBA equation \eqref{eq:cTBA-0} but with the boundary
condition at $\theta \rightarrow -\infty$ given by
\begin{equation}
  \label{zamobc}\epsilon (\theta,P) \sim 8 P \theta-2C(P), \qquad \theta \to
   -\infty, \quad P>0,
 \end{equation}
where $C(P)$ is written down in (\ref{CP}). 
Let us now introduce the function 
\be X(\mu, P)=\exp\left[-\epsilon(\theta, P)/2\right], 
\ee
where $\mu$ is related to $\theta$ by
\be
\label{mutheta}
\mu=  \re^{2 \theta}\left(\frac{\Gamma \left(\frac{1}{4}\right)^2}{16 \sqrt{\pi }}\right)^2. 
\ee
Then, according to \cite{post-zamo}, the spectral determinant of the modified Mathieu operator (\ref{eq:schrodinger}) is given by 
\be  
\Xi (E)= {X(\mu, P) \over X(\mu, 0)},
\ee
where the parameters $\Lambda$, $E$ and $\hbar$ of the operator are related to the parameters appearing in $X(\mu, P)$ by
\be \label{diczo} \mu=  \Lambda^2 \hbar^{-2}, \qquad P^2 =-E \hbar^{-2}.
 \ee 
   If we compare the result of \cite{post-zamo} with ours we should have (by using the dictionary \eqref{diczo}) 
 \be {X(\mu, P) \over X(\mu, 0)}=A (\hbar)\left( \sinh\left(\frac{ \Pi_A^{\rm ex}(E, \hbar)}{2 \hbar }\right) \right)^{-1} \cosh\left({1 \over 2\hbar} \Pi_B^{\rm ex}(E, \hbar)\right). 
 \ee
 In order to find the relation between the two normalization constants
 $X(\mu, 0)$ and $A (\hbar)$, it is useful to first derive the
 asymptotic behavior \eqref{zamobc} from our expression \eqref{sd}. We
 need to expand around small $\Lambda \hbar^{-1}$ and take $u<0$,
 which means that $a$ is imaginary, as discussed in (\ref{aminus}). In
 this regime, and by using \eqref{abpe}, we have
 \be 
 \ba
 &\label{asy} \left( \sinh\left(\frac{\pi a}{\hbar }\right) \right)^{-1} \cosh \left({\ri \over \hbar}\partial_{a}F_{\rm NS}(a, \hbar)\right)\\
 & \approx  2^{-1}\pi^{-1}\left(\left({\Lambda\over\hbar}\right)^{-2s }\Gamma(1+s)\Gamma(s)+\left({\Lambda\over\hbar}\right)^{2s }\Gamma(1-s)\Gamma(-s)\right),
 \ea
 \ee
 where 
 \be s=-\ri {a \over \hbar} = \frac{\alpha}{\hbar} > 0.
 \ee
By using \eqref{mmap} we have 
\be 
E \approx a^2/4 
\ee
and therefore 
%
 \be P =s/2 >0.
 \ee
Hence we can neglect the second term in the r.h.s. of \eqref{asy}. It follows from \eqref{diczo} that \eqref{asy} agrees precisely with \eqref{zamobc}. 
In particular this means that the two normalization constants are identified and we have
\be \label{IDtest} X(\mu, P)= \left( \sinh\left(\frac{ \Pi_A^{\rm ex}(E, \hbar)}{2 \hbar }\right) \right)^{-1} \cosh\left({1 \over 2\hbar} \Pi_B^{\rm ex}(E, \hbar)\right). 
  \ee
We test this equality by solving numerically the TBA equation \eqref{eq:cTBA-0} with the boundary condition \eqref{zamobc}. 
Some results  are given in Table \ref{TestTBA1}. We find perfect agreement.

 \begin{table}[h!] 
\centering
   \begin{tabular}{l l l}
  \\
 $N_b$& $\log\left(\left( \sinh\left(\frac{\pi a}{\hbar }\right) \right)^{-1} \cosh\left({\ri \over \hbar}\partial_{a}F_{\rm NS}(a, \hbar)\right)\right)$  \\
\hline
 2 &\underline{11.360025}317439438\\
 4 &\underline{11.36002529911}2863\\ 
  6&11.360025299117259\\  
 \hline
TBA &  11.360025299117    \\
\end{tabular}    \\
\caption{ The un-normalized spectral determinant as computed 
by using instanton counting and 
by solving numerically  the TBA \eqref{eq:cTBA-0} with \eqref{zamobc}.  We use $P^2=-E=5$, $\mu=\Lambda^2=\left(\Gamma \left(1/4\right)^2/(16 \sqrt{\pi })\right)^2$, and $\hbar=1$. The number $N_b$ means that we truncate the series \eqref{inst}  at order $\Lambda^{4 N_b}$. }
 \label{TestTBA1}
 \end{table}

 An important spinoff of this comparison is that our result (\ref{sd})
 provides an analytic, closed form solution to the TBA equation of
 \cite{post-zamo}. This also has the following consequence.  When we
 derived the Fredholm determinant from the topological string
 perspective, and due to our regularization procedure, we generated an
 integration constant $A(\hbar)$ whose explicit expression is given in
 \eqref{ainst}.  Given the identity \eqref{IDtest} between our
 Fredholm determinant and the solution to the Zamolodchikov's TBA, we
 expect $A(\hbar)$ to be computed by the integral equation \eqref{eq:cTBA-0} at
 $P=u=0$.  More precisely we expect
\be \label{ccc} 2 \log A(\hbar) = \epsilon(\theta, P=0),
\ee
where we used the dictionary \eqref{diczo}.  For $P=0$ the asymptotic
condition \eqref{zamobc} does not make sense, strictly speaking.
Nevertheless, we can derive the appropriate asymptotic condition for
the TBA at $P=0$ by using our analytic expression \eqref{IDtest}.  We
find that, as $\theta \rightarrow -\infty$,
%
%
\begin{equation}\label{eps0}
  \epsilon(\theta, P=0) 
  \sim -2 \log \left(
    -\frac{2 (\theta +\gamma_{\rm Euler} )-\log \pi+4 \log\Gamma(5/4)}{\pi}
  \right).
\end{equation}
 This is precisely the boundary condition used in section~\ref{tbau0}, equation \eqref{eq:bdy-cor}. 
 One can now check \eqref{ccc} numerically.
  For instance, by solving the TBA of  section~\ref{tbau0} we find
 \be \epsilon(\theta,P=0)\Big|_{\theta=-1}=0.51888\cdots 
 \ee
 Likewise, by using instanton counting, and in particular \eqref{abpe} and \eqref{ainst}, we have ($\Lambda=1$)
 \be 2 \log\left(A \left(\frac{16 \re \sqrt{\pi }}{\Gamma \left(\frac{1}{4}\right)^2}\right)\right)=0.51887965286656 \cdots
 \ee
We have 5 matching digits which is consistent with the precision achieved with the TBA equation.

This discussion provides an additional result along the lines of what we obtained in section \ref{bo}. As we discussed in section 
\ref{tbau0}, the function $\epsilon(\theta)$ with the boundary condition (\ref{eq:bdy-cor}) computes the dyonic  period $\Pi_D(0,\hbar)$. As pointed out in Sec \ref{bo}, 
such period is Borel summable, and we can indeed test that its Borel resummation agrees with \eqref{ainst}, 
namely
\be  \label{bspd}\exp\left({1 \over 2 \hbar} s\left(\Pi_D\right)(0,\hbar)\right)={\sinh\left(\frac{\pi a}{\hbar }\right) \over \cosh\left({\ri \over \hbar}\partial_{a}F_{\rm NS}(a, \hbar)\right)}\bigg|_{a=a(0, \hbar)}=\frac{\sinh\left(\frac{1}{2 \hbar}  \Pi_A^{\rm ex}(0, \hbar)\right) }{ \cosh\left({1 \over 2\hbar} \Pi_B^{\rm ex}(0, \hbar)\right)}.
\ee
We have verified this identity numerically. In addition we have tested that \eqref{bspd} also holds for other values of $u$ in the strong coupling region, and we 
conjecture that, for $u\in [-1, 1]$, one has 
\be
 \label{bspd2}\exp\left({1 \over 2 \hbar} s\left(\Pi_D\right)(u,\hbar)\right)=\frac{\sinh\left(\frac{1}{2 \hbar}  \Pi_A^{\rm ex}(u, \hbar)\right) }{ \cosh\left({1 \over 2\hbar} \Pi_B^{\rm ex}(u, \hbar)\right)}.
\ee
%

\section{On the modified Mathieu operator and Painlev\'e $\rm III_3$}
\label{mathieu-painleve}
It was observed by many authors \cite{Lukyanov:2011wd, PCL, nok} that
the movable poles of Painlev\'e $\rm III_3$ are somehow related to
Mathieu functions. In particular in \cite{ sciarappa2}, based on
\cite{bgt}, it was observed that the zeros of the Painlev\'e
$\rm III_3$ $\tau$ function compute the spectrum of modified Mathieu
(with a suitable dictionary).  From the view point of the TS/ST
correspondence \cite{ghm} this connection comes naturally since both
systems arise as limiting cases of this duality.  In particular, the
modified Mathieu operator arises in the standard geometric engineering
limit \cite{hm,gm17,gm3}, while Painlev\'e $\rm III_3$ arises in the
dual geometric engineering limit considered in \cite{bgt}.

In this section we prove the connection between the zeros of the Painlev\'e $\rm III_3$ $\tau$ function and the spectrum of the modified Mathieu operator by using \cite{ggu}. 
From the CFT perspective this is a connection between Liouville conformal blocks at $c=1$ and $c=\infty$. We proceed as follows. 
First we  write the Painlev\'e $\rm III_3$ $\tau$ function as \cite{gil,ilt,Gavrylenko:2017lqz}
\be
 \label{GIL} \tau(\Lambda, a, \eta, \hbar)= \sum_{n \in \IZ}  \re^{4 \pi \ri n \eta} \exp\left(F^{\rm SD}(a+2\ri\hbar n, \hbar, \Lambda)\right)
 \ee
where
\be
\ba  \exp\left(F^{\rm SD}(a, \hbar, \Lambda)\right)= &\left(\frac{\Lambda }{\hbar }\right)^{-\frac{  a^2}{\hbar ^2}} {1 \over  G(1-\ri \frac{a}{\hbar })G(1+\ri \frac{a}{\hbar })}\\
&\times \left(1 -\frac{2\Lambda ^4}{ a^2 \hbar ^2}+\frac{\Lambda ^8 \left(2 a^2 -\hbar ^2\right)}{ a^2 \hbar ^2 \left( a^2+\hbar ^2\right)^2}+\mathcal{O}(\Lambda^{12})\right)\ea
\ee
is the so-called four dimensional Nekrasov partition function in the selfdual $\Omega$ background \cite{n} (namely, the equivariant parameters are $\epsilon_1=-\epsilon_2=\hbar$).
The parameters $(a, \eta)$ in \eqref{GIL} play the role of initial conditions while $\Lambda$ is the time.
We are interested in the case in which 
\be \eta=0.
\ee  
We now recall the result of
\cite{ggu}, where it was demonstrated that, in the NS limit, the
Nakajima--Yoshioka blowup equations for $SU(2)$ pure SYM
\cite{ny1,naga,nagalect} can be written as\footnote{Strictly speaking this
  is the four-dimensional limit of \cite{ggu}. This type of
  expressions first appeared in \cite{swh} as compatibility conditions
  between the exact quantization conditions of \cite{ghm,cgm} and
  those of \cite{wzh,fhm,hm}. A different connection between blowup and Painlev\'e equations was used in \cite{bes, Bershtein:2018zcz,ntalkbu} to prove the so-called Kiev formula \cite{gil,ilte} or its q-deformed version \cite{bsu, jns-qp}.
  }
\be\label{comp}
		\sum_{n\in\mathbb{Z} }
		\exp\left(\ri n  \pi+{F}^{\rm SD}\left(a+ 2\ri\hbar n+\ri\hbar, \hbar, \Lambda\right)
		 -2 \ri n\hbar^{-1}\frac{\partial }{\partial a}{F}_{\text{NS}}\left(a,\hbar\right)\right)=0.
\ee
Finally, we use the quantization condition for the modified Mathieu operator in the NS form (\ref{eqref}). It then follows from \eqref{comp} that, if a value of 
$a$ satisfies this exact quantization condition, one finds a vanishing condition for the tau function of
Painlev\'e $\rm III_3$, namely
\be\label{tau0}
 \tau\left(\Lambda, a+{\ri \hbar}, 0, \hbar \right)=\sum_{n\in\mathbb{Z} }
		\exp\left({F}^{\rm SD}\left(a+ 2\ri\hbar n+\ri\hbar, \hbar, \Lambda\right) \right)=0.\ee
Notice that we think of \eqref{eqref} and \eqref{tau0} as quantization conditions for the variable $a$. In order to obtain the spectrum of modified Mathieu one has to use the quantum Matone relation \eqref{mmap}.

\section{Conclusions}\label{concl}

In this paper we have used non-perturbative techniques inspired by
supersymmetric gauge theory and topological string theory to study the
quantization of the Seiberg--Witten curve of $\CN=2$, $SU(2)$ super
Yang--Mills theory, which gives the modified Mathieu operator. On the
one hand, building upon \cite{gmn, gmn2, oper, ims}, we have obtained
integral equations for the Borel resummation of the quantum periods
obtained with the all-orders WKB method. These equations predict as
well the resurgent structure of these periods, and in particular their
Stokes discontinuities. The results obtained in this way have been
tested against calculations in the WKB method to very high order. We
have also clarified the relation between these Borel-resummed quantum
periods and the ``exact" quantum periods given by instanton calculus
(in the NS limit). On the other hand, we have used the TS/ST
correspondence of \cite{ghm,cgm} to obtain a closed formula for the
spectral determinant of the modified Mathieu operator, and we have
compared this formula to previous results by Zamolodchikov.

Our results raise several issues. An important problem concerns the
relation between the TBA equations obtained in the context of SW
theory, and the analytic bootstrap program first proposed in
\cite{voros-quartic} and reloaded in \cite{ims,mm-s2019}. In the TBA
equations obtained in \cite{oper,ims} for quantum mechanics with
polynomial potentials, one only needs the boundary condition
associated to the classical behavior (i.e.\ at $\hbar \rightarrow 0$,
or equivalently at $\theta \rightarrow \infty$). The boundary behavior
when $\theta \rightarrow -\infty$ is fixed by the integral
equations. As pointed out already in \cite{oper} and further discussed
in section \ref{tbau0} of this paper, the integral equations for the
modified Mathieu operator admit many possible boundary conditions at
$\theta \rightarrow -\infty$, and one needs additional information to
fix them. One can use the quantum Matone relation and the first
quantum correction to the periods to obtain additional constraints.
However, it seems clear from the study of this example that the
analytic bootstrap might require additional asymptotic information to
determine uniquely the resummed quantum periods. As suggested in
\cite{oper}, one might obtain the appropriate boundary conditions by
first solving the full TBA equations of \cite{gmn} (before taking the
conformal limit) and then implementing the conformal limit directly on
the solution.

Another problem that should be discussed more carefully is how to
solve efficiently the TBA equations to compute the Borel resummed
quantum periods. In particular, we should understand in detail how to
to solve the infinite tower of TBA equations appearing in the weak
coupling region.

It would be very interesting to extend the techniques developed in
this paper to quantum mirror curves. This would provide a relation
between BPS states in local CY threefolds (studied for example in
\cite{dfr}) and the resurgent properties of the corresponding quantum
periods. Work along this direction has been already done in
\cite{eager, longhi}.  Another interesting class of quantum curves
which could be studied with our methods is the one given by quantum
A-polynomials of knots (see e.g. \cite{exact-dimofte}). In this case,
the resurgent properties should be closely related to the resurgent
properties of Andersen--Kashaev invariants \cite{ak}, which have been
considered in \cite{gmp, gh,andersen}. They might correspond to BPS
states in the supersymmetric dual obtained with the 3d/3d
correspondence of \cite{dgg}.

Another intriguing point is the following. Based on previous works
\cite{oper,ims}, we have shown that the conformal limit of the GMN TBA
equations encode in a precise way the NS limit of the Omega background
for the pure SU(2) theory. On the other hand it is interesting to
observe that, as pointed out in \cite{bgt}, there is another set of
TBA equations which computes the selfdual limit of the Omega
background. The latter was obtained by Zamolodchikov in \cite{zamo},
see also \cite{Fendley:1999zd}.  Interestingly also such TBA can be
obtained from \cite{gmn} upon a suitable limiting procedure.  It would
be interesting to investigate more concretely if and how the full TBA
equations of \cite{gmn} encode the full Omega background. Work along
this direction was performed in \cite{Cecotti:2014wea}.
	
In addition it should be possible to extend the results of section
\ref{mathieu-painleve} to Painlev\'e $\rm III_2, III_1, V$ and
$\rm VI$. In these cases, the r\^ole of the modified Mathieu operator
is replaced by the quantum SW curve of $SU(2)$ gauge theory with
$N_f=1,2,3,4$ flavours, respectively. In particular, for $N_f=4$ one
should recover the connection between Painlev\'e $\rm VI$ and the Heun
operator \cite{Litvinov:2013sxa} (see also
\cite{Lencses:2017dgf}). Likewise the spectrum of the Calogero-Moser
system should make contact with the $\tau$ function describing the
isomonodromic deformations on the torus \cite{Bonelli:2019boe}. The
details will appear somewhere else \cite{BGG-p}.

The situation for Painlev\'e $\rm I, II, IV$ is more subtle since
these correspond to Argyres--Douglas theories of type $H_0, H_1, H_2$,
respectively \cite{blmst}.  At present we do not know how to write
Nakajima-Yoshioka blowup equations for these theories.  Nevertheless,
it should be possible to connect the NS limit to the selfdual limit of
the $\Omega$ background also in these theories, since the $H_i$
theories can be derived from $SU(2)$ gauge theories with $N_f=1,2,3$
upon a suitable limiting procedure \cite{ad,Argyres:1995xn}. By
following \cite{gg-qm,ito-shu}, such connection would provide a
relation between the exact spectrum of the quantum SW curves
underlying the $H_0, H_1, H_2$ theories, and Painlev\'e
$\rm I, II, IV$ tau functions.  Note that the quantum SW curve of the
$H_0$ and $H_1$ theories correspond to the cubic and quartic
oscillators, respectively. Connections between Painlev\'e equations
and the above quantum mechanical systems have been observed in
\cite{dm1,dm2,nok2,bender-painleve}.

\section*{Acknowledgements}
We would like to thank Vladimir Bazhanov, Katsushi Ito, Davide
Gaiotto, Qianyu Hao, Lotte Hollands, Amir-Kian Kashani--Poor, Sergei
Lukyanov, Davide Masoero, Gregory Moore, Andy Neitzke, Stefano Negro,
Nikita Nekrasov, Hongfei Shu, Leon Takhtajan, Roberto Tateo and Peter
Wittwer for useful discussions and correspondence.
The work of J.G.\ and M.M.\ is supported in part by the Fonds National
Suisse, subsidy 200021-175539 and by the NCCR 51NF40-182902 ``The
Mathematics of Physics'' (SwissMAP).

\appendix

\section{The four dimensional $SU(N)$ spectral determinant}
\label{asun}

In this Appendix we explain how the exact formula (\ref{sd}) can be
extended to the family of operators studied in \cite{gm3}. These operators have the form, 
\be
\label{q-ham}
\mH_N=\Lambda^N\left(  \re^\mm+\re^{-\mm} \right) + \sum_{k=0}^{N-1} (-1)^k \mx^{N-k} h_k, \quad [\mx,\mm]=\ri \hbar,
\ee
where $N\ge 2$ is a positive integer and we set $h_0=1$, $h_1=0$. They can be regarded as deformations of the standard non-relativistic Schr\"odinger operators with a 
polynomial potential. When $N$ is even, they have a discrete spectrum and their inverses are of trace class. When $N$ is odd, one can 
perform a standard analytic continuation and obtain a discrete spectrum of resonances, as explained in \cite{gm3}. In both cases, one can define 
a Fredholm determinant as  
\be
 \Xi_N(h_2, \cdots, h_N)= \det\left(1+{h_N (-1)^N \over \mH_N} \right). 
 \ee
 Here, $h_2, \cdots, h_{N-1}$ are the moduli appearing in the potential, while $(-1)^{N-1} h_N$ can be identified with the energy and is the standard auxiliary variable appearing in 
 the definition of Fredholm determinants. 
 
As explained in \cite{gm3}, we can engineer the following operator
from the quantum mirror curve to the $Y^{N,0}$ geometry. 
We follow \cite{gm3} and define
\be
\label{spec-vector}
\bs{\gamma}= {1\over 2}\sum_{i=1}^N (-1)^{i-1} \bs{e}_i, 
\ee
where $\bs{e}_i$ are the weights 
of the fundamental representation of  $SU(N)$. We denote by
\be
\label{orbit}
\CW_N\cdot \bs{\gamma} =\left\{ w(\bs{\gamma}): w\in \CW_{N} \right\}
\ee
the Weyl orbit of $\bs{\gamma}$, and  we introduce
\be \bs{a} = \sum_{j=1}^{N-1}a_i\bs{\lambda},
\ee
where $\{ \blam_i\}_{i=1, \cdots, N-1}$ are the fundamental weights of $SU(N)$.
The quantities $a_i$ are related to the parameters $h_i$ in \eqref{q-ham} by using the four dimensional mirror maps 
or quantum Matone relations (see  for instance eqs (3.95)--(3.107) in \cite{gm3} and reference therein). 
For instance, we have
\be
\label{qmm-ex}
\ba
h_2\left(\bs{a}; \hbar \right)&=\lim_{\epsilon_2 \to 0}\left( -{1\over Z} \sum_{\bs{Y}} 
\left((-1)^N \Lambda^{2N} \right)^{\ell (\boldsymbol{Y})}  \CC_2 (\bs{a}, \bs{Y}) \CZ_{\boldsymbol{Y}}\right),\\
h_3\left(\bs{a};  \hbar \right)&=\lim_{\epsilon_2 \to 0}\left( {2\over Z} \sum_{\bs{Y}}  \left((-1)^N \Lambda^{2N} \right)^{\ell (\boldsymbol{Y})}  \CC_3 (\bs{a}, \bs{Y}) \CZ_{\boldsymbol{Y}}\right), 
\ea
\ee
where $\CZ_{\boldsymbol{Y}}, Z$ are defined in \eqref{zdef} and \eqref{z4d}, and $\bs{Y}$ is a vector of Young diagrams as in \eqref{Y-vec}.
Moreover,
\be
\label{ccs}
\ba
\CC_2(\bs{a}, \bs{Y}) &={1\over 2} \sum_{I=1}^N \alpha_I^2 -\ri \hbar \epsilon_2 \ell (\boldsymbol{Y}) , \\
\CC_3(\bs{a}, \bs{Y}) &= \ri \hbar \epsilon_2 \left( {\ri \hbar +\epsilon_2 \over 2} \ell (\boldsymbol{Y}) +\ri \hbar 
\sum_{I=1}^N c_2(Y^t_I) +\epsilon_2 
\sum_{I=1}^N c_2(Y_I) - \sum_{I=1}^N \alpha_I \ell(Y_I) \right) \\
&+ {1\over 6} \sum_{I=1}^N \alpha_I^3, 
\ea
\ee
where $ \ell (\boldsymbol{Y})$ is defined in \eqref{ly} and we use
 \be
 c_2(Y)= {1\over 2} \sum_{i\ge1}y_i (y_i-1). 
\ee
We also denote
\be a_i=\alpha_{i}-\alpha_{i+1} , \quad i=1,\cdots, N-1,
\ee
with
 \be  \sum_{i=1}^N \alpha_i=0. \ee
With a procedure analogous to the one of section~\ref{deri}, 
we obtain the explicit formula
\be
\Xi_N(h_2, \cdots, h_N)= A_N(\hbar, \Lambda, { h_2}, \cdots h_{N-1})\sum _{{\bf n} \in \CW_N\cdot \bs{\gamma}} \re^{J_{\bs n}^{\rm 4d}}.
\ee
 The quantity $\re^{J_{\bs n}^{\rm 4d}}$ is defined as follows. If $N$ is even we have:
\be 
\re^{J_{\bs n}^{\rm 4d}}= \exp\left( {\ri \over \hbar} {\partial F_{\rm NS} \over 
\partial \bs{a}}\cdot \bs{n} \right) \prod_{\alpha \in \Delta_+} 
\left( 2 \sinh \left( {\pi \bs{a}\cdot \balpha\over \hbar}\right) \right)^{-\left(\bs{n}\cdot \balpha\right)^2},
\ee
while if $N$ is odd we have
\be 
\re^{J_{\bs n}^{\rm 4d}}= 
\exp\left( {\ri \over \hbar} {\partial F_{\rm NS} \over \partial \bs{a}}\cdot \bs{n} -{\pi \over \hbar}\bs{a} \cdot \bs{n} \right) \prod_{\alpha \in \Delta_+} \left( 2 \sinh \left( {\pi \bs{a}\cdot \balpha\over \hbar}\right) \right)^{-\left(\bs{n}\cdot \balpha\right)^2}, 
\ee
where $F_{\rm NS}$ is defined in \eqref{fns4d}. The quantity $A_N(\hbar, \Lambda, { h_2}, \cdots h_{N-1})$ is an integration constant, analogous to $A(\hbar)$ in (\ref{sd}), which 
now depends on the moduli $h_2, \cdots, h_{N-1}$. The above spectral determinant vanishes precisely when the quantization conditions 
obtained in \cite{gm3} are satisfied. When $N=2$ we recover exactly the result \eqref{sd}. When $N=3$ we have
\be
\label{sd3}
\ba 
\Xi_3 (h_3,h_2 ) &=A_3(\hbar, \Lambda, h_2)\left[1+ {1 -\re^{-2 \pi a_1/\hbar} \over 1- \re^{-2 \pi (a_1+a_2) /\hbar}}\re^{-2 \pi a_2/\hbar} \re^{\ri \phi_2} + 
{1 -\re^{-2 \pi a_2/\hbar} \over 1- \re^{-2 \pi (a_1+a_2) /\hbar}}  \re^{\ri \phi_1}\right]\\
&\times \text{csch}\left(\frac{\pi  a_1}{\hbar }\right) \text{csch}\left(\frac{\pi  a_2}{\hbar }\right) \re^{-\frac{\pi  (a_1-a_2)}{3 \hbar }-\ri {1\over 3}  (\phi_1+\phi_2)},
\ea
\ee
where $\phi_i$, $i=1,2$, are defined as \cite{gm3}
\be
 \phi_1(a_1, a_2; \hbar)={1\over \hbar} \left( {\partial F_{\rm NS} \over \partial a_2} - 2  {\partial F_{\rm NS} \over \partial a_1} \right), \qquad
 \phi_2(a_1, a_2; \hbar)= {1\over \hbar} \left(
2 {\partial F_{\rm NS} \over \partial a_2} -  {\partial F_{\rm NS} \over \partial a_1} \right). 
\ee
We have tested \eqref{sd3} by expanding the r.h.s~of \eqref{sd3} around $h_3=0$ and comparing with the numerical values of the spectral traces. We  find perfect agreement.

\section{Zamolodchikov's TBA equation for the modified Mathieu
   equation}\label{appendixzamo}

In \cite{zamo-sg} Zamolodchikov considered the thermodynamic TBA
ansatz for the sinh-Gordon model. This model depends on the
parameter $b\in\IC$, and we introduce
\begin{equation}
  Q=b+{1\over b}, 
\end{equation}
as well as 
\begin{equation}
  \label{parameters}
  p= {b^2\over 1+ b^2}, \qquad  a=1-2p={1-b^2 \over 1+b^2}. 
\end{equation}
The TBA equation for this theory is given by 
\begin{equation}
\label{fullshgTBA}
  \epsilon (\theta)=m R \cosh(\theta)-\left( \phi\star L \right)(\theta), 
\end{equation}
In this equation, $R$ is the radius of the circle where the theory lives, $m$ is the mass of the particle in the spectrum,
\begin{equation}
  L(\theta) = \log\left(1+ \re^{-\epsilon(\theta)}\right),
\end{equation}
and
\begin{equation}
\label{sh-kernel}
  \phi(\theta)= {1\over 2 \pi} \left( {1\over\cosh(\theta-\ri \pi
      a/2)} +  {1\over\cosh(\theta+\ri \pi a/2)}  \right)= {1\over 2
    \pi} {4 \sin (\pi p) \cosh(\theta) \over  
    \cosh(2 \theta) - \cos(2 \pi p)}. 
\end{equation}
The $\star$ in (\ref{fullshgTBA}) denotes, as it is standard, the convolution
\be
\left(f\star g \right)(\theta)= \int_\IR f(\theta-\theta') g(\theta') \rd \theta'. 
\ee
The ground state energy is then given by
\begin{equation}
  E(R)= -{m \over 2 \pi} \int_\IR \cosh(\theta) L(\theta) \rd \theta, 
\end{equation}
and the effective central charge is 
\begin{equation}
  c_{\rm eff}=-{6 R \over \pi} E(R). 
\end{equation}

The formal conformal limit of the above TBA equation was analyzed in
\cite{post-zamo} in relation to the generalized Mathieu equation
\begin{equation}
  -u''(x)+ \left(\mu_- \re^{-bx}+ \mu_+\re^{bx}\right) u(x)=-P^2 u(x). 
\end{equation}
The parameters $\mu_\pm$ have the following obvious symmetry
\begin{equation}
  \mu_+ \to \mu_+ \re^{-\varepsilon/b}\ ,\quad \mu_-
  \to\mu_-\re^{+\varepsilon b} \ ,\quad x\to x+\varepsilon \ ,
\end{equation}
and therefore only the combination
\begin{equation}
  \mu = \mu_+^b \mu_-^{1/b}
\end{equation}
matters. The parameter $b$ is identified with the parameter of the sinh-Gordon model, $\mu$ corresponds to its coupling constant, while the energy
\begin{equation}
  E=-P^2
\end{equation}
is identified with the Liouville momentum, and enters into the effective central charge of the theory, see (\ref{ceff}). 
In the conformal limit, the TBA equation (\ref{fullshgTBA}) 
becomes
\begin{equation}\label{eq:TBA-gen}
  \epsilon (\theta)=\pi \re^\theta-2 \left(\phi\star L \right)(\theta). 
\end{equation}
The dependence on $P$ comes through as the boundary condition of the
TBA solution when $\theta\rightarrow -\infty$,
\begin{equation}
  \label{bcP}
  \epsilon(\theta) \sim 4 QP \theta-2 C(P)+\cdots
\end{equation}
where $P>0$ and
\begin{equation}
\label{CP}
  C(P) = \log\frac{\Gamma(2P)\Gamma(1+2P)}{2\pi} +
  4P\log\frac{16\sqrt{\pi}}{\Gamma(1/4)^2} \ .
\end{equation}
It is then argued in \cite{zamo-sg} that the Fredholm determinant of the generalized
Mathieu equation is given by \cite{post-zamo}
\begin{equation}
  \Xi(\mu,P) = {X(\mu,P) \over X(\mu,0)},
\end{equation}
where
\begin{equation}
  X(\mu,P) = \exp\[-\epsilon(\theta,P)/2\]
\end{equation}
and $\mu$ is related to $\theta$ by (\ref{mutheta}). We have indicated the explicit dependence of $\epsilon$ on $P$ through the boundary condition (\ref{bcP}).

The ordinary modified Mathieu equation is obtained when 
\begin{equation}
b=1, \qquad \mu_-=\mu_+=\mu. 
\end{equation}
Let us focus on this case. The TBA equation becomes
\begin{equation}
  \label{zmathieu}
  \epsilon (\theta)=\pi \re^{\theta}-\int_\IR {L(\theta') \over \cosh(\theta-\theta')} {\rd \theta' \over \pi}. 
\end{equation}
To impose the boundary condition (\ref{bcP}), we use a trick due to Zamolodchikov. We first note that, 
as a consequence of (\ref{bcP}), we have
\begin{equation}
  L(\theta) \sim -8 P \theta +2 C(P), \qquad \theta \rightarrow -\infty, 
\end{equation}
and we introduce the function
\begin{equation}
  L_0(\theta)= 4 P \log(1+ \re^{-2 \theta}), 
\end{equation}
which has the same leading asymptotics than $L(\theta)$, 
\begin{equation}
  L_0(\theta) \sim -8 P \theta + \CO(\re^{-|\theta|}), \qquad
  \theta\rightarrow -\infty.   
\end{equation}
We have
\begin{equation}
f_0=2\phi \star L_0 =8 P \log(1+ \re^{-\theta}), 
\end{equation}
and we can rewrite the TBA equation as 
\begin{equation}
\epsilon (\theta)= \pi \re^\theta -f_0 -2\phi\star(L-L_0). 
\end{equation}
This has by construction the right asymptotic behavior (\ref{bcP}). 

One property of (\ref{zmathieu}) which is relevant for our analysis is
the following. The asymptotic behavior of the solution
$\epsilon(\theta)$ as $\theta \rightarrow \infty$ is of the form
\begin{equation}
\epsilon(\theta)= \pi \re^{\theta} +\epsilon^{(1)} \re^{-\theta}+ \cdots, 
\end{equation}
where 
\begin{equation}
\epsilon^{(1)}=-{2 \over \pi} \int_\IR \re^\theta \log\left(1+\re^{-\epsilon(\theta)}\right) \rd \theta. 
\end{equation}
On the other hand, this correction is proportional to the effective
central charge of the theory\footnote{There is a factor of $2$ missing
  in eq. (4.4) of \cite{post-zamo}.},
\begin{equation}
c_{\rm eff}= {6 \over  \pi} \int_\IR \re^\theta \log\left(1+\re^{-\epsilon(\theta)}\right) \rd \theta=-3 \epsilon^{(1)} ,
\end{equation}
which according to \cite{post-zamo} can be computed in terms of $P$ only
\begin{equation}
\label{ceff}
c_{\rm eff}= 1+ 24 P^2. 
\end{equation}
This means that
\begin{equation}
\label{fqc}
\epsilon^{(1)}=-{1\over 3}\left(1+ 24 P^2 \right). 
\end{equation}

\bibliographystyle{JHEP}

\linespread{0.6}
\bibliography{biblio}

\end{document}